%% file: turbulent_reactivity.tex
\documentclass[%
 reprint,
nofootinbib,
amsmath,amssymb,
 aip,
floatfix,
longbibliography,
]{revtex4-2}

\usepackage{graphicx}% Include figure files
\usepackage{subcaption}
\usepackage{dcolumn}% Align table columns on decimal point
\usepackage{bm}% bold math
\usepackage[english]{babel}
\usepackage{upgreek}
\usepackage{ifthen}
\usepackage{comment}
\usepackage{tikz}
\usepackage{pgfplots}
\pgfplotsset{compat=1.17}

\usepackage{diagbox}

\input{macros.tex}

\newcommand{\uh}{\hat{u}}

\newcommand{\prel}{p_\mr{rel}}
\newcommand{\ptot}{p_\mr{tot}}

\newcommand{\pctwo}{p_\mr{c2}}
\newcommand{\pczero}{p_\mr{c0}}
\newcommand{\pctworel}{p_\mr{c2,rel}}
\newcommand{\pctwotot}{p_\mr{c2,tot}}

\newcommand{\pcone}{p_\mr{c1}}
\newcommand{\pconerel}{p_\mr{c1,rel}}

\newcommand{\dflog}{h}

\newcommand{\citeFF}{\cite{Fetsch_Fisch_2024}}
\newcommand{\FF}{FF}

\newcommand{\reactop}{\Sigma}

\usepackage{mathptmx}
\usepackage{etoolbox}

\makeatletter
\def\@email#1#2{%
 \endgroup
 \patchcmd{\titleblock@produce}
  {\frontmatter@RRAPformat}
  {\frontmatter@RRAPformat{\produce@RRAP{*#1\href{mailto:#2}{#2}}}\frontmatter@RRAPformat}
  {}{}
}%
\makeatother
\begin{document}

%\preprint{APS/123-QED}
\preprint{AIP/123-QED}

\title{Analytical models for the enhancement of fusion reactivity by turbulence}

\author{Henry Fetsch}
 \email{hfetsch@princeton.edu}
\author{Nathaniel J. Fisch}%
\affiliation{Department of Astrophysical Sciences, Princeton University, Princeton, NJ}

%TODO: refer to Lavine et al IEEE trans plasma sci 2024. z pinch special issue. TKE and complete energy balance in z pinch

\date{\today}

%COAUTHOR QUESTIONS
% Title...
% Abbreviation (SFRE etc...)

\begin{abstract}
The reactivity of fusion plasma depends not only on its local density and temperature but also, through a recently identified kinetic effect, on the relative velocities of nearby fluid elements. 
Turbulence on fine spatial scales therefore enhances fusion reactivity. 
The enhancement is quantified here for general subsonic turbulent flows. 
Leveraging this effect in the design of inertial confinement fusion (ICF) experiments could enable substantial energy savings.
\end{abstract}

\maketitle

%TODO (general)
%mention that flow puts kinetic energy mainly into ions, whereas heating generally entails equipartition between species (especially relevant for Z > 1)

\section{Introduction}
\label{sec_intro}

Fusion reactivity is typically assumed to depend only on the temperature of the reacting ions \cite{Lindl_1995,Rosen_2024,Betti_et_2010}. It is thus widely believed that maximizing reactivity in inertial confinement fusion (ICF) requires directing as much of the implosion kinetic energy as possible into ion heating. According to this view, any turbulent kinetic energy (TKE) remaining at bang time is wasted, leading to lower fusion gain.

In fact, this is not necessarily the case. 
In regions of strong shear, the kinetic behavior of fast ions increases fusion reactivity far above its thermal value. This ``shear flow reactivity enhancement effect'' (\sfe) was recently identified by Fetsch and Fisch (\FF)\citeFF. 
A remarkable consequence of the \sfe~is that turbulence, in which many swirling eddies moving relative to each other lead to flow shear throughout the system, actually enhances fusion reactivity. 
The implications for ICF are profound: rather than waiting for all of the implosion kinetic energy to thermalize, TKE might, in fact, be put to advantage. 
While the \sfe~ is easily understood in qualitative terms and in simplified geometries\citeFF, no general quantitative description of the effect in turbulent plasma currently exists. 
The present work addresses this gap, providing analytical formulas for the \sfe~in general subsonic turbulent flows.

Suppression of turbulence generally takes high priority in the design of ICF experiments \cite{Kritcher_et_2024,Hurricane_et_2024,Rosen_2024}. 
Nevertheless, both magnetically driven \cite{Giuliani_et_2014,Kroupp_et_2011,Maron_2020} and laser-driven implosions \cite{Clark_et_2015,Kritcher_et_2024,Zhou_Sadler_Hurricane_2025} frequently exhibit high levels of TKE. 
In fuel near the regime of thermonuclear instability -- the ``ignition cliff'' -- fusion yield is extremely sensitive to the efficiency of compression and burn \cite{Lindl_Haan_Landen_Christopherson_Betti_2018,Clark_Haan_Salmonson_2008,Betti_et_2010,Christopherson_Betti_Lindl_2019}, meaning that even a small change reactivity can have a large effect on implosion performance. 
The intriguing possibility suggested by the \sfe~is that, even in current ICF experiments, the fusion yield may be improved by designing a target to direct the already extant TKE into smaller-scale eddies. 
Deliberately driving additional TKE at the expense of heating comes with some advantages, including reduced conductive and radiative losses \cite{Davidovits_Fisch_2016,Davidovits_Fisch_2018,Davidovits_Fisch_2019}.

As a further benefit, cold fuel stops alpha particles in a shorter distance. Therefore, if fuel can be made cooler because a portion of reactivity is provided by the \sfe, ignition becomes possible in a smaller hot spot. The ability to ignite a smaller fuel mass allows for dramatic energy savings \citeFF. 
The effect is particularly notable in fast ignition, where FF showed that the combination of lower temperature and smaller hot spot volume can reduce the ignitor energy requirement by an order of magnitude. 
Therefore, substantial improvements in gain might be realized by optimizing ICF designs to take advantage of the \sfe. Such optimization requires a quantitative theory of the reactivity enhancement in turbulent plasma, which this work provides.

%The \sfe~represents a unique feature of turbulence in fusion plasmas, granting it some intrinsic academic interest. The effect may appear in magnetic confinement fusion (MCF) and astrophysical plasmas as well. It is in ICF, however, where the \sfe~appears likely to have dramatic consequences. 

%Accounting for the \sfe~ opens new opportunities for ICF designs deriving an advantage from turbulence. 
%Reducing heating in favor of small-scale TKE offers a reduction in conductive and radiative losses. 
%Ignition at lower temperature, made possible when reactivity is enhanced by the \sfe, shortens the stopping distance of charged fusion products, allowing ignition in a smaller hot spot. The ability to ignite a smaller fuel mass leads to remarkable energy savings, particularly in fast ignition\citeFF. 

The results presented in the following sections quantify the reactivity enhancement with increasing levels of accuracy but with increasing complexity.
First, \S\ref{sec_asymp} describes the limiting behavior of the \sfe~in turbulence on small length scales. The result, \eqref{eq_Phi_gk_infty_low_Mach}, captures the general dependence of the reactivity enhancement on temperature and flow velocity. 
Next, \S\ref{sec_analytics} derives formulas valid in more general turbulent flows. In \eqref{eq_G_def}, the reactivity enhancement is shown to take a simple form, depending only on the turbulent energy spectrum $E(k)$ and a general ``utility function'' $G(k)$. 
An asymptotically accurate formula for $G(k)$ is derived in \eqref{eq_G_k_asymptotic}; while useful for estimation, it breaks down at high temperatures. A corrected formula, \eqref{eq_G_k_corrected}, is valid at ICF-relevant temperatures and constitutes the central result of this work. 
In \S\ref{sec_numerics}, the analytical results are validated using numerical simulations. 
Finally, in \S\ref{sec_discussion}, the \sfe~ is evaluated under conditions relevant to ICF and shown to be significant when turbulence is driven on small scales.
To set the stage for the main derivations, \S\ref{sec_setup} first reviews some key physics relevant to fusion reactivity.

\section{Fusion reactivity}
\label{sec_setup}

%\subsection{Thermal and Gamow scales}
%\label{subsec_setup_kinetics}

In homogeneous plasma, fusion reactivity is simply a function of the ion temperature (the overall fusion rate is the product of the reactivity and the number densities of the reactants). Even in inhomogeneous plasma, reactivity is typically evaluated as a function of the local temperature at each point. 
Previous works have identified nonlocal corrections to reactivity; a well-known example is the Knudsen-layer reactivity reduction \cite{Molvig_et_2012,Albright_et_2013} accounting for the depletion of fast ions, whose long mean free paths lead them to escape ICF hot spots at an increased rate. 
Such corrections, including the \sfe \citeFF, are hints of kinetic physics appearing in systems otherwise well described by hydrodynamics. 
%According to conventional estimates, kinetic effects are usually negligible in the burn phase of ICF
%Conventional estimates would suggest that kinetic effects can safely be ignored in most ICF systems, but this is not always the case. 
%In fact, kinetic effects induce an order-unity correction to fusion reactivity. Ignoring the nonlocal effect of shear flow on reactivity can lead to dramatic underestimation of the fusion rate.

Kinetic effects on fusion reactivity owe their surprising relevance to the scale separation between the particles that determine viscosity and those that undergo fusion. 
In a near-equilibrium distribution, most ions have velocities near the thermal velocity $\vth \doteq \sqrt{T/m}$, where $T$ is the temperature\footnote{Temperature has units of energy in this work.} and $m$ is the ion mass. 
By weight of numbers, these thermal ions determine bulk fluid properties, such as temperature and viscosity. 
Spatial inhomogeneities cause a departure from local thermodynamic equilibrium, whose size is described by the Knudsen number $\kn \doteq \lambda_\mr{th}/L$. Here, $\lambda_\mr{th} \doteq \vth/\nu_0$ is the mean free path of thermal ions, $\nu_0$ is the characteristic collision frequency, and $L$ is a characteristic lengthscale over which fluid properties vary. 
When $\kn \ll 1$, gradients are considered weak, the distribution function $f$ remains close to a Maxwellian, and corrections to $f$ can be computed using standard Chapman-Enskog theory. 
Such systems are well described by hydrodynamics.

This picture breaks down for fast particles. 
The ion-ion collision frequency $\nu_\mr{ii}$ depends strongly on velocity $v$, scaling as $\nu_\mr{ii} \sim (v/\vth)^{-3}\nu_0$ for fast ions ($v \gg \vth$). The mean free path for ion-ion collisions therefore scales as $\lambda_\mr{ii} \sim (v/\vth)^4\lth$. 
The total ion collision frequency is $\nu = \nu_\mr{ii} + \nu_\mr{ie}$, where $\nu_\mr{ie}$ is the ion-electron collision frequency. 
Although $\nu_\mr{ie}$ is small, scaling as the square root of the electron-ion mass ratio $\sqrt{m_e/m}$, it depends weakly on $v$ (provided that the ion velocity is much less than the electron thermal velocity $v_\mr{th,e}$), so slowing down on electrons is the dominant collision process for fast ions. 
The mean free path can therefore be approximated by 
\begin{equation}
    \label{eq_lambda_approx}
    \lambda \sim \frac{v^4}{\nu_0} \frac{1}{1 + v^3\sqrt{m_e/m}}
\end{equation}
for ions with $\vth \ll v \ll v_\mr{th,e}$. Because the characteristic velocity of ions participating in fusion reactions is typically several times greater than the thermal velocity, the mean free path of these ions, according to \eqref{eq_lambda_approx}, can be orders of magnitude longer than $\lambda_\mr{th}$. Thus, even when $\kn \ll 1$, meaning that thermal ions explore only small regions within one mean free path and sample nearly uniform fluid properties, fast ions can travel long distances and cross large differentials in fluid properties.

%\subsection{Fusion reactivity}
%\label{subsec_setup_reactivity}

The importance of fast ions for fusion reactions stems from the fact that these particles are more capable than thermal particles of tunneling through the Coulomb barriers of other ions. This tendency is reflected in the fusion cross section $\sigma$. For nonresonant reactions between ions of a single species, as for deuterium-deuterium (DD) fusion, $\sigma$ can be written as \cite{Bosch_Hale_1992}
\begin{equation}
    \label{eq_sigma_S}
    \sigma(\prel) = \frac{S(\prel)}{\vth^2 \prel^2} e^{-b/\prel} ,
\end{equation}
where $\prel \doteq (\bs w - \bs w')/\sqrt{2} \vth$ is the normalized relative velocity of reacting ions whose velocities are $\bs w$ and $\bs w'$, $S$ is a rational function, and $b$ is defined by
\begin{equation}
    \label{eq_b_def}
    b \doteq \frac{\sqrt{2E_G}}{\vth},
\end{equation}
where $E_G$ is the Gamow energy. Typically, $b \gg 1$ in regimes relevant to laboratory fusion. For DD fusion, $b \approx 44.4 T^{-1/2}$, where $T$ has units of keV.

In this work, we describe fusion reactivity for same-species reactions using the bilinear reactivity operator $\reactop$ defined as 
\begin{equation}
    \label{eq_reactivity_def}
    \reactop[f_a, f_b] \doteq \frac{\vth}{\sqrt{2}} \iint d^3 w d^3 w' f_a(\bs w) f_b (\bs w') \prel \sigma(\prel) 
\end{equation}
so that the total fusion reactivity $\avg{\sigma v}$ of a plasma with ion distribution function\footnote{Distribution functions are normalized to one in this work and do not carry a factor of overall number density.} $f$ is $\avg{\sigma v} \doteq \reactop[f, f]$. 
For the special case of Maxwellian reactant distributions $f_M$ and a cross section given by \eqref{eq_sigma_S}, the reactivity can be simplified to
\begin{equation}
    \label{eq_reactivity_int_maxwellian}
    \reactop[f_M, f_M] = \frac{1}{(2\pi)^{3/2} \sqrt{2} \vth}\int d^3 \prel \frac{S(\prel)}{\prel} e^{-b/\prel - \half \prel^2} .
\end{equation}

The rational prefactor in \eqref{eq_reactivity_int_maxwellian} varies slowly relative to the exponential, which has a maximum at ${\prel = p_* \doteq b^{1/3}}$, the ``Gamow peak." Treating $b$ as an asymptotically large parameter, $\reactop$ can be approximated analytically by Laplace's method, expanding the integrand about $p_*$ to find
\begin{equation}
    \label{eq_reactivity_maxwellian_leading}
    \reactop[f_M, f_M] \sim \sqrt{\frac{2}{3}} \frac{b^{1/3} S(b^{1/3})}{\vth} e^{-\frac{3}{2}b^{2/3}} .
\end{equation}

The characteristic velocity of ions involved in fusion reactions is therefore $v_* \doteq p_*\vth$, which is typically much greater than the thermal velocity. The corresponding ``Gamow mean free path'' is $\lambda_*\doteq \lambda(v_*)$. 
As a result, the dimensionless number capturing the importance of kinetic effects for fusion reactions is not the Knudsen number but rather the ``Gamow-Knudsen number''\cite{Molvig_et_2012,McDevitt_Tang_Guo_2017,Fetsch_Fisch_2024} $\gk \doteq \lambda(v_*)/L$  . 
For large $b$, \eqref{eq_lambda_approx} guarantees that $\gk \gg \kn$; heuristically, $\mr{Gk} \sim (v_*/\vth)^4\mr{Kn}$. Fusion-relevant ions ``see'' much longer length scales than thermal ions and are sensitive to inhomogeneities that their thermal cousins would fail to ``notice.''
%The Knudsen and Gamow-Knudsen numbers are related by $\gk \sim p_*^4 \kn$, and so the scales that they describe can easily differ by orders of magnitude. 

This scale separation is central to this work, justifying the separate consideration of thermal ions (with hydrodynamics) and fusion-relevant suprathermal (with kinetic theory). It furthermore helps to explain why the \sfe, like related effects such as the Knudsen-layer reactivity reduction, were long neglected in studies of ICF: an estimation of $\kn$ may conclude that kinetic effects are insignificant, but, if applied naively, this process leads to the erroneous neglect of fast-ion dynamics and sometimes to significant mis-estimation of reactivity.

We denote the kinetic enhancement (or reduction) in reactivity by $\Phi$, defined as
\begin{equation}
    \label{eq_Phi_def}
    \Phi[f_a, f_b] \doteq \frac{\reactop[f_a, f_b]}{\reactop[f_M, f_M]} ,
\end{equation}
where $f_M$ is a Maxwellian matching the (number density-weighted) average zeroth and second moments of $f_a$ and $f_b$. 
It is convenient to isolate the nonmaxwellian behavior in a factor $\varphi$ such that
\begin{equation}
    f_\mr{a}(\bs p)f_\mr{b}(\bs p') = \varphi_\mr{ab}(\bs p, \bs p') f_M(p)f_M(p') .
\end{equation}

If $\varphi_\mr{ab}$ varies slowly %in $\bs p$ and $\bs p'$ 
compared to the exponential in \eqref{eq_reactivity_int_maxwellian}, then $\Phi$ can be approximated as 
\begin{equation}
    \label{eq_Phi_approx_phi}
    \Phi[f_a, f_b] \sim \left<\varphi_\mr{ab}\paren{\frac{b^{1/3}}{\sqrt{2}}\hat r,-\frac{b^{1/3}}{\sqrt{2}}\hat r} \right>_{\Omega},
\end{equation}
where $\hat r$ is a vector on the unit sphere and $\avg{\cdot}_\Omega$ denotes an average over the unit sphere. 
Note that \eqref{eq_Phi_approx_phi} is only accurate to leading order in $b$; its simple form is a consequence of the fact that the Gamow peak does not shift significantly as long as the exponential remains the most rapidly varying part of the integrand. 
The remainder of this work takes a series of approaches to determine $\Phi$ with increasing fidelity.

\section{Limiting behavior}
\label{sec_asymp}

We begin by determining bounds on $\Phi$. To isolate the \sfe, we consider here only systems with uniform density and temperature. The flow field is unspecified except for the requirement that it only vary on length scales consistent with $\kn \ll 1$. 
The assumption of uniform density requires that the flow be solenoidal. 
Since $\gk$ can take a range of values, we here consider its limits.  
When $\gk \rightarrow 0$, the forcing away from local thermodynamic equilibrium vanishes and so $f \rightarrow f_M$. Then, by definition, ${\Phi \rightarrow 1}$.

%--\textcolor{blue}{TODO:} entropy argument for $\Phi \geq 1$ given our assumptions?

When $\gk \rightarrow \infty$, the mean free path of ions near the Gamow peak is much longer than the gradient scale length $L$ of the flow. Suppose that $\bs u (\bs x)$ is a turbulent flow field and is sufficiently random that any straight line of sufficient length averages over all values of $\bs u$ realized in the system. Effectively, then, an ion with velocity $v_* \gg \vth$ can be treated as colliding not with the local Maxwellian, but with an averaged background distribution ${F(\bs v) \doteq \avg{f_M (\bs w = \bs v - \bs u(\bs x))}_\lambda }$, where $f_M$ is stationary in the local fluid frame and $\avg{\cdot}_\lambda$ denotes the average along a trajectory of length $\lambda \gg L$.

In this case, the fast-ion distribution $f$ can be expected to come to equilibrium with $F$. 
This equilibrium should be a Maxwellian with some effective temperature $T_\mr{eff}$ determined by the second moment of $F$, \textit{viz.}
\begin{equation}
    T_\mr{eff} = m\vth^2 + \frac{m\avg{u^2}}{3}
\end{equation}
where $m$ is the mass of the ion species.
In other words, $T_\mr{eff}$ can be viewed as the sum of thermal temperature and the ``turbulent temperature.'' Let $\tau \doteq T_\mr{eff}/T$. 
To find $\reactop[f, f]$, we can simply use the expression already derived for the reactivity of a Maxwellian \eqref{eq_reactivity_maxwellian_leading}, with $\vth$ replaced by $\vth \tau^{1/2}$ and with $b$ replaced by $b \tau^{-1/2}$. To leading order in $b$, the reactivity enhancement in the $\gk \to \infty$ limit therefore approaches
\begin{equation}
    \label{eq_Phi_gk_infty}
    \Phi \to \frac{1}{\tau^{2/3}} \frac{S\paren{\frac{b^{1/3}}{\tau^{1/6}}}}{S(b^{1/3})} e^{\frac{3}{2}b^{2/3}\paren{1 - \tau^{-1/3}}} .
\end{equation}

This expression simplifies considerably in the low Mach-number limit. When $\avg{u^2} \ll b^{-1/3}\vth^2$, a subsidiary expansion yields, to leading nontrivial order in the Mach number, 
\begin{equation}
    \label{eq_Phi_gk_infty_low_Mach}
    \Phi \to 1 + \frac{1}{6} \frac{\avg{u^2} b^{2/3}}{\vth^2} .
\end{equation}

The asymptotics of $\Phi$ at small and large Gamow-Knudsen number, which were obtained here under few assumptions, provide useful grounding for the analytical and numerical results in the rest of this work.

\section{Analytical results}
\label{sec_analytics}

This section develops an analytical theory for the reactivity of unmagnetized plasmas with arbitrary flow fields. 
For the sake of tractability, we use a simplified collision operator in the following calculation and restrict our consideration to low Mach-number flows. Formally the problem can be posed as follows: given a plasma with a flow field $\bs u(\bs x)$ and thermal velocity $\vth$, and whose ions fuse with a cross section described by \eqref{eq_sigma_S}, what is the reactivity if $b \gg 1$, $\kn \ll 1$, and $|u| \ll \vth$ everywhere?

\subsection{Description of the system}
\label{subsec_analytics_system}

The ion distribution function $f(\bs x, \bs v)$, written in terms of position $\bs x$ and velocity $\bs v$, evolves according to the kinetic equation
\begin{equation}
    \label{eq_vlasov_v}
    \partial_t f + \bs v \cdot \nabla f + \bs a \cdot \frac{\partial f}{\partial \bs v} = \mc C [f] ,
\end{equation}
where $\bs a$ is the acceleration due to electromagnetic and gravitational forces, and $\mc C$ is the collision operator. We consider a single species of mass $m$ in thermal equilibrium at temperature $T$. It will frequently be convenient to work with the peculiar velocity (the velocity in the local rest frame of the fluid) $\bs w \doteq \bs v - \bs u$ or with the normalized peculiar velocity $\bs p \doteq \bs w/\vth$. Locally, the equilibrium distribution is the Maxwellian
\begin{equation}
    \label{eq_maxwell}
    f_M(\bs p) = \frac{1}{(2\pi)^{3/2}} e^{-\half p^2} 
\end{equation}
(magnitudes of vectors are marked in this work by standard typeface, e.g. $p = |\bs p|$, and components are marked by subscripts, e.g. $\bs p = (p_x, p_y, p_z)$). In terms of peculiar velocity, \eqref{eq_vlasov_v} reads
\begin{equation}
    \label{eq_vlasov_w}
    \partial_t f + (\bs w + \bs u) \cdot \nabla f + \paren{\bs a - \frac{d \bs u}{dt}- \bs w \cdot \nabla \bs u}\cdot \frac{\partial f}{\partial \bs w} = \mc C[f] ,
\end{equation}
where $d \bs u/dt$ includes acceleration due to hydrodynamic forces, such as pressure gradients and viscosity.

We consider here a solenoidal flow field $\bs u(\bs x)$ with low Mach number, meaning respectively that ${\nabla \cdot \bs u = 0}$ and ${|\bs u| \ll \vth}$ for all $\bs x$. The Knudsen number is assumed to be small so that the bulk of the distribution is close to a Maxwellian. For a simplified collision operator, we adopt a modified Bhatnagar-Gross-Krook (BGK) operator defined by 
\begin{equation}
    \label{eq_bgk_def}
    \mc C[f] = -\nu(w) (f - f_M),
\end{equation}
where $\nu(w)$ is a collision rate that depends on the particle velocity relative to the local Maxwellian. This operator is unsuitable for a general calculation because it does not conserve fluid number, momentum, or energy. In this work, however, we are interested only in suprathermal particles, which have large $\lambda$. This population is small enough relative to the thermal population to have a negligible effect on the fluid moments. To conserve important quantities, a small correction can be applied to the final solution.

The system under consideration is stationary in time with no external forces, and $\kn$ is taken to be small enough to neglect viscous forces. With these approximations, \eqref{eq_vlasov_w} reduces to
\begin{equation}
    \label{eq_kinetic_analytical_approx}
    \bs w \cdot \nabla f - \bs w \cdot \nabla \bs u \cdot \partial_{\bs w} f = - \nu(w) (f - f_M) .
\end{equation}

Determining corrections to reactivity requires computing the non-Maxwellian tail of the distribution function that solves \eqref{eq_kinetic_analytical_approx}. 
It would be conventional to take gradients to be weak and expand in the Knudsen number, but because we are interested in fast ions, whose mean free paths could be comparable to the gradient scale length, this is not an option. 
Instead, we proceed by an expansion in the Mach number. To begin, we decompose $\bs u (\bs x)$ into its Fourier components as follows:
\begin{equation}
    \label{eq_u_fourier_def}
    \wt {\bs u} (\bs k) \doteq \frac{1}{\vth \mc V} \int d^3x e^{i \bs k \cdot \bs x} \bs u (\bs x) 
\end{equation}
where $\mc V$ is the system volume times $(2\pi)^3$. We define a small parameter $\uh \ll 1$ and consider flow fields in which $\wt u \sim \mc O(\uh)$ for all Fourier components. The normalized energy spectrum is defined as
\begin{equation}
    \label{eq_energy_spectrum_def}
    E(k) \doteq \half \int d^3k' |\wt {\bs u}(\bs k')|^2 \delta(k' - |\bs k|) 
\end{equation}
such that the total TKE is given by
\begin{equation}
    \label{eq_TKE_spec}
    \mr{TKE} = T \int_0^\infty dk \, E(k) .
\end{equation}

Although ${\kn \ll 1}$, the quantity 
\begin{equation}
    \label{eq_gamma_def}
    \gamma(k, w) \doteq \frac{k w}{\nu(w)} = k\lambda
\end{equation}
is not necessarily small for fast particles. In other words, the following calculation applies even in regimes where the Gamow-Knudsen number is of order unity and where a large reactivity enhancement can be expected.
%In the expansion that follows, it is assumed that $\gamma \sim \mc O(1)$ in the velocity regime of interest. 
%For convenience, we define the normalized peculiar velocity $\bs p \doteq \bs w /\vth$.

\subsection{Perturbed distributions}
\label{subsec_pert_dist}

\subsubsection{First-order distribution}

We proceed to calculate $f$ order by order in $\uh$, denoting the nth-order correction by $f_n$ and its spatial Fourier transform by $\wt f_n$. At zeroth order, $f_0 = f_M$. To first order, each Fourier mode may be considered separately. Without loss of generality, let us align our coordinates for the mode of wavenumber $\bs k$ such that $\bs k = k \bs e_z$ and $\wt{\bs u}(\bs k) = \wt u \bs e_x$, where $\bs e_x$ and $\bs e_z$ are unit vectors in the $x$ and $z$ directions, so \eqref{eq_kinetic_analytical_approx} becomes
\begin{equation}
    \label{eq_vlasov_f1}
    -i k p_z \wt f_1(\bs k, \bs p) + i k p_z \wt u \frac{\partial f_0}{\partial p_x} = - \frac{\nu (p)}{\vth} \wt f_1(\bs k, \bs p) .
\end{equation}
Defining $\chi \doteq p_x/p$ and $\xi \doteq p_z/p$, the first-order distribution function corresponding to this Fourier mode is
\begin{equation}
    \label{eq_f1_kp}
    \wt f_1(\bs k, \bs p) = \wt u(\bs k) f_0 \frac{i \gamma(k, p) \xi \chi p }{1 - i\gamma(k, p) \xi}.
\end{equation}

\subsubsection{Second-order distribution}

It is necessary to compute $f$ to second order in $\uh$ because the average of $f_1$ over space vanishes -- indeed, the average of $f$ over angles also vanishes at each point -- so ${\reactop[f_1, f_0] = 0}$. 
Finding $f_2$ is cumbersome in general, but, conveniently, the only term that survives spatial averaging to contribute to $\reactop[f_2, f_0]$ is the $\bs k=0$ Fourier component. This greatly simplifies the calculation because, while $f_2$ is sourced by the nonlinear interaction of $f_1$ and $\bs u$, $\wt f_2(0, \bs p)$ is sourced exclusively by interactions between modes of the flow and modes of $f_1$ with wavenumbers that sum to zero (e.g. $\wt{\bs u}(\bs k)$ and $\wt f_1(-\bs k)$). 

Therefore, without loss of generality, we can begin by considering a flow field with Fourier components only at some $\bs k$ and $-\bs k$. The coordinate system is aligned with $\bs k$ as above. Denoting the zero-wavenumber Fourier component corresponding to this mode by $\wt f_2(0, \bs p; \bs k)$, the kinetic equation yields at second order
%For this field, we have
%\begin{equation}
 %   \label{eq_f2_kp_start}
  %  \wt f_2(0, \bs p; \bs k) = -\frac{\vth p_z}{\nu(p)} \left[-i k \wt u(\bs k) \frac{\partial \wt f_1(-\bs k)}{\partial p_x} + ik \wt u(-\bs k) \frac{\partial \wt f_1(\bs k)}{\partial p_x}\right] .
%\end{equation}
%Because $\bs u(x)$ and $f_1(\bs x, \bs p)$ are both real, \eqref{eq_f2_kp_start} can be simplified to 
\begin{equation}
    \wt f_2(0, \bs p; \bs k) = \xi \gamma(k, p) 2 \Re \left[i \wt u(\bs k) \frac{\partial \wt f_1^*(\bs k)}{\partial p_x}\right] ,
\end{equation}
where $\Re[\cdot]$ indicates the real part of the argument. Defining for convenience
\begin{equation}
    \label{eq_g_def}
    g(p) \doteq p \frac{\partial \ln \gamma(k, p)}{\partial p}
\end{equation} 
and using \eqref{eq_f1_kp}, we find (dropping $\bs k$ and $\bs p$ arguments where there is no ambiguity)
\begin{equation}
    \label{eq_f2_kp}
    \wt f_2(0, \bs p; \bs k) = 2 |\wt u|^2 f_0 \frac{\gamma^2 \xi^2}{1 + \gamma^2 \xi^2}\left[p^2 \chi^2 - 1 - \frac{g \chi^2 (1 - \gamma^2 \xi^2)}{1 + \gamma^2 \xi^2}\right] .
\end{equation}

At this point, we can average over angles and find $\wt f_2$ as a function of $p$ because, to second order in $\uh$, the only contribution to reactivity from $f_2$ will come from collisions with an isotropic Maxwellian. Denoting by $\avg{\cdot}_\Omega$ the average over the unit sphere, we have
\begin{equation}
    \label{eq_f2_avg} 
    \begin{split}
    \avg{\wt f_2(0, \bs p; \bs k)}_{\Omega} =  \frac{1}{2}|\wt u|^2 f_0 &\Bigg[p^2 \frac{4\gamma^3 - 6(\gamma^2 + 1)\tan^{-1}(\gamma) + 6\gamma}{3\gamma^3} 
    \\ & - 4\paren{1 - \frac{\tan^{-1}(\gamma)}{\gamma}} 
    \\ &- g \frac{12(\gamma^2 + 2)\tan^{-1}(\gamma) - 4\gamma(\gamma^2 + 6)}{3\gamma^3} \Bigg] .
    \end{split}
\end{equation}

%Formally, the expansion used in deriving \eqref{eq_f2_avg} has assumed that $\gamma \uh p \ll 1$, and so is only valid at velocities for which this ordering holds. However, the solutions retain some value outside of this regime; \eqref{eq_f1_kp} and \eqref{eq_f2_avg} remain asymptotically self-consistent provided that $\uh p \ll 1$.

%\subsubsection{Limiting cases}

It is useful to consider the limits of \eqref{eq_f2_avg} for small and large Gamow-Knudsen number. Holding $p$ and $\wt u$ constant while varying $\gamma$, we find 
\begin{equation}
    \label{eq_f2_gamma_0}
    \avg{\wt f_2(0, \bs p; \bs k)}_{\Omega} \xrightarrow{\gamma \to 0} \frac{1}{2}|\wt u|^2 f_0 \left[\paren{\frac{4}{15}p^2 - \frac{4}{3} - \frac{4}{15}g }\gamma^2 + \mc O(\gamma^4)\right]
\end{equation}
and 
\begin{equation}
    \label{eq_f2_gamma_infty}
    \avg{\wt f_2(0, \bs p; \bs k)}_{\Omega} \xrightarrow{\gamma \to \infty} \frac{1}{2}|\wt u|^2 f_0 \left[\frac{4}{3}p^2 - 4 - \frac{4}{3}g + \mc O\paren{\frac{1}{\gamma}}\right] .
\end{equation}
For $p \sim \mc O(p_*)$, \eqref{eq_f2_gamma_0} and \eqref{eq_f2_gamma_infty} correspond to the limits of very small and very large Gamow-Knudsen number respectively.

It bears emphasizing that these perturbed distribution functions are valid for fast particles only; the bulk is assumed to be nearly Maxwellian. 
Strictly speaking, the tail enhancement found here is accompanied by a depletion of the bulk. However, in the limit where $\kn \ll 1$, this depletion is small; formally, we neglect all terms below leading order in $\kn$. This is fully consistent with a large reactivity enhancement; exchanging a small number of particles between the bulk and the tail has a much larger effect, proportionally speaking, on the tail.

\begin{comment}
For a general flow field described by \eqref{eq_u_fourier_def}, the second-order distribution function is given by
\begin{equation}
    \label{eq_f2_avg_integral}
    \avg{f_2(p)}_{\Omega, \mc V} = \int d^3k \half |\wt u(\bs k)|^2 \avg{\wt f_2(0, p; \bs k)}_{\Omega} ,
\end{equation}
where $\avg{\cdot}_{\Omega, \mc V}$ denotes an average over the unit sphere and the system volume (the factor of $\half$ appears here to avoid double counting because \eqref{eq_f2_avg} corresponds to a pair of modes with opposite wavenumbers). %and $\avg{\wt f_w(0, p; \bs k)}_\Omega$ is given by \eqref{eq_f2_avg} for a mode with wavenumber $\bs k$.
\end{comment}

\subsection{Reactivity enhancement}
\label{subsec_analytics_reactivity}

To second order in $\uh$, the reactivity is
\begin{equation}
    \label{eq_reactivity_expansion_second_order}
    \reactop[f, f] \sim \reactop[f_0, f_0] + \reactop[f_1, f_1] + 2\reactop[f_2, f_0] .
\end{equation}
By definition, $\reactop[f_0, f_0] = 1$. The latter two terms are of second order in $\hat u$. 
%Our problem therefore reduces to finding effective expressions for the final two terms of \eqref{eq_reactivity_expansion_second_order}. 
%In general, no closed-form solutions exist. For some applications, reactivity may be evaluated numerically using \eqref{eq_f1_kp} and \eqref{eq_f2_avg} as a starting point; this is addressed in \S\ref{sec_numerics}.
Conveniently, to this order, the reactivity enhancement takes the following form:
\begin{equation}
    \label{eq_G_def}
    \avg{\Phi[f, f]}_V \sim 1 + 2\int_0^\infty dk E(k) G(k) ,
\end{equation}
where $\avg{\cdot}_V$ denotes an average over space. The utility function $G(k)$ describes the degree of reactivity enhancement associated with energy in modes of scale $k$ and is given by 
\begin{widetext}
\begin{equation}
    \label{eq_G_fraction}
    G(k) = \left\langle\frac{\int d^3 p \int d^3 p' \frac{S(\prel)}{\prel} e^{-b/\prel - \half p^2 - \half {p'}^2} \left[\varphi_{11}(\bs p, \bs p'; \bs k) + 2 \varphi_{20}(\bs p,\bs p';\bs k) \right]}{\int d^3p \int d^3p' \frac{S(\prel)}{\prel} e^{-b/\prel - \half p^2 - \half {p'}^2}} \right\rangle_\Omega.
\end{equation}
\end{widetext}
The remainder of this section derives analytical formulas for $G(k)$, which, while necessarily approximate, are useful for quickly estimating the reactivity enhancement.

\subsubsection{Reactivity integrands}

\begin{sloppypar}
%The first term of \eqref{eq_reactivity_expansion_second_order} is simply the Maxwellian reactivity, for which $\Phi = 1$ by definition. 
%For the remaining terms of \eqref{eq_reactivity_expansion_second_order}, 
%The nonmaxwellian parts of the reactivity integrands, as defined in \S~\ref{sec_setup}, can be written as ${\varphi_{20} (\bs p, \bs p') = \int d^3 k \varphi(\bs p, \bs p'; \bs k)}$, where
The second nonmaxwellian term in the numerator of \eqref{eq_G_fraction} is simply\footnote{The factor of 2 appears in \eqref{eq_phi20} to avoid double counting because \eqref{eq_f2_avg} corresponds to a pair of modes with opposite wavenumbers.}
\begin{equation}
    \label{eq_phi20}
    \varphi_{20}(\bs p, \bs p'; \bs k) = \frac{\wt f_2(0, \bs p; \bs k)}{2f_M(p)} .
\end{equation} 
For concision, we can define ${\overline{\varphi}_{20}(p; k) \doteq \avg{\varphi_{20}(\bs p, \bs p'; \bs k)}_\Omega}$ because neither $\bs p'$ nor the direction of $\bs p$ will be relevant. 
To compute the first nonmaxwellian term, we have
\begin{equation}
    \label{eq_phi11}
    \varphi_{11}(\bs p, \bs p'; \bs k) = \frac{\Re \left[\wt f_1(\bs k, \bs p)\wt f^*_1(\bs k, \bs p')\right]}{f_M(p)f_M(p')}  ,
\end{equation}
where the real part is taken for convenience because the imaginary part will vanish when all $\bs k$ are added together. Using \eqref{eq_f1_kp}, we find
\begin{equation}
    \label{eq_phi11_angles}
    \begin{split}
    \varphi_{11}(\bs p, \bs p', \bs k) = & |\wt u(\bs k)|^2 \gamma\gamma' \xi \xi' \chi\chi' p p' 
    \\
    &\times \left[\frac{1 + \gamma \gamma' \xi \xi' }{(1 + \gamma^2 \xi^2)(1 + {\gamma'}^2 {\xi'}^2)} \right]
    \end{split}
\end{equation}
where $\chi' \doteq p_x'/p'$ and $\chi' \doteq p_z'/p'$ and, for concision, $\gamma = \gamma(k, p)$ and $\gamma' = \gamma(k, p')$. 
We will be particularly interested in nearly head-on collisions, for which $\bs p \approx -\bs p'$. In this case, \eqref{eq_phi11_angles} simplifies considerably and can be averaged over angles to obtain
\begin{equation}
    \label{eq_phi11_angles_avg}
    \begin{split}
    \overline{\varphi}_{11}(p; k) = &\frac{1}{4}|\wt u(\bs k)|^2 p^2
    \\&\times\frac{12(\gamma^2 + 2)\tan^{-1}(\gamma) - 4\gamma(\gamma^2 + 6)}{3\gamma^3}   ,
    \end{split}
\end{equation}
where we defined for concision ${\overline{\varphi}_{11}(p; k) \doteq \avg{\varphi_{11}(\bs p, -\bs p; \bs k)}_\Omega}$.
\end{sloppypar}

\subsubsection{Asymptotic reactivity formula}

In the limit where $b \to \infty$, \eqref{eq_phi20} and \eqref{eq_phi11_angles} can be used to find an asymptotic formula for $\Phi$ by expanding the integrand about the Gamow peak located at ${\prel = p_* = b^{1/3}}$ with ${\bs p' = -\bs p}$. This procedure, formalized in \eqref{eq_Phi_approx_phi}, is valid as long as the exponential in \eqref{eq_G_fraction} remains the most rapidly varying part of the integrand. The Gamow-Knudsen number is ${\gamma_* \doteq \gamma(k, b^{1/3}/\sqrt{2})}$. It follows immediately from \eqref{eq_f2_avg} and \eqref{eq_phi11_angles_avg} that
%\begin{equation}
 %   \label{eq_G_k_corrected}
  %  \Phi[f, f] \sim 1 + \int d^3 k \frac{1}{4} |\wt u(\bs k)|^2 \frac{b^{2/3}}{2}\frac{12\tan^{-1}(\gamma_*) + 4\gamma_*^3 - 12\gamma_*}{3\gamma_*}
%\end{equation}
\begin{equation}
    \label{eq_G_k_asymptotic}
    G(k) \sim \frac{b^{2/3}}{2}\paren{\frac{1}{3} + \frac{\tan^{-1}(\gamma_*) - \gamma_*}{\gamma_*^3}} ,
\end{equation}
where only the term proportional to $p^2$ is retained because $b^{2/3} \gg 1$ in the limit under consideration. Note that \eqref{eq_G_k_asymptotic} reduces in the $\gamma_* \to \infty$ limit to the result \eqref{eq_Phi_gk_infty_low_Mach} obtained in \S\ref{sec_asymp} by simpler arguments. 
Formally, the combination of \eqref{eq_G_k_asymptotic} and \eqref{eq_G_def} constitutes a solution to the problem posed at the start of this section, but it fails to describe DD fusion under ICF-relevant conditions.

\subsubsection{Corrected reactivity formula}

In reality, while $b$ is generally large enough to justify the assumption of a separation between $\kn$ and $\gk$, it is often not large enough for a naive application of Laplace's method in evaluating \eqref{eq_reactivity_def}. Consider, for example, $b = 27$, as is the case near $3~\text{keV}$ for DD fusion. Suppose that $\gamma_* \ll 1$ for all relevant modes and, for simplicity, let $\gamma \propto p^4$. Then, by \eqref{eq_f2_gamma_0}, ${\overline \varphi_{20} \propto p^{10}}$. Noting that $b/p_* = 9$, it is clear that the exponential part of the reactivity integrand no longer varies much faster than $\varphi$ and so \eqref{eq_Phi_approx_phi} no longer applies. Physically, this means that the Gamow peak shifts to larger velocity. Here, we derive a formula \eqref{eq_G_k_corrected} accounting for this shift. 

The Gamow peak shifts differently for $\reactop[f_2, f_0]$ and for $\reactop[f_1,f_1]$. For the former, we define parameters $\pctwo$ and $\pczero$ such that, at the shifted peak, the particle sampled from $f_2$ has velocity $p = \pctwo \hat r$ (where $\hat r$ is an arbitrary direction) and the particle sampled from $f_0$ has velocity $\bs p' = -\pczero \hat r$. For collisions between two particles sampled from $f_1$, we define $\pcone$ such that colliding particles at the Gamow peak have velocities $\bs p = \pcone \hat r$ and $\bs p' = -\pcone \hat r$.

To estimate the shifts, we approximate $f_2$ and $f_1$ by power laws in the region between the old and new Gamow peaks. There is some freedom in approximating the exponents, but the following choices are used here:
\begin{eqnarray}
    \label{eq_dflog2_def}
    \dflog_2 &\doteq& 2 + b^{2/3} \frac{d}{dp}\paren{\ln\overline{\varphi}_{20;p^2}(p; k)}  \Big|_{p=b^{2/3}}, \\
    \label{eq_dflog1_def}
    \dflog_1 &\doteq& b^{2/3} \frac{d}{dp}\paren{\ln\overline{\varphi}_{11}(p; k) } \Big|_{p=b^{2/3}} .,
\end{eqnarray}
so that the scalings $\overline \varphi_{20} \propto p^{\dflog_2}$ and $\overline \varphi_{11} \propto p^{\dflog_1}$ hold approximately in the region of interest. Here, $\overline \varphi_{20;p^2}$ means only the terms of $\overline \varphi_{20}$ that are proportional to $p^2$ (the first line of \eqref{eq_f2_avg}); this helps to isolate the power-law behavior of $\overline \varphi_{20}$, which dominates at high velocity.

When $\dflog_2$ and $\dflog_1$ are comparable to $b^{2/3}$, the peak shifts significantly. For collisions between $f_2$ and $f_0$, we group the rapidly varying parts of the integrand into the term ${\exp(-\chi_2(\prel, \ptot; \bs k))}$, where the quantity
\begin{equation}
    \label{eq_chi2_def}
    \chi_2(\prel, \ptot; \bs k) \doteq \frac{\sqrt{2}b}{\prel} + \frac{\prel^2}{2} + \frac{\ptot^2}{2} - \ln \varphi_{20}(\prel, \ptot; \bs k) 
\end{equation}
is maximized at the shifted Gamow peak. Let ${\pctworel \doteq (\pctwo - \pczero)/\sqrt{2}}$ and ${\pctwotot \doteq (\pctwo + \pczero)/\sqrt{2}}$ be the relative and center-of-mass velocities at the shifted Gamow peak. Their approximate values are determined by the following formulas, which can be viewed as the first steps of an iterative solution to \eqref{eq_chi2_def}. The relative velocity at the peak $\pctworel$ is defined to be equal to the largest real root of the polynomial
\begin{equation}
    \label{eq_p2rel_soln}
    y_2(s) \doteq s^4 - \dflog_2 s^2 - b s + (s^3 - b)\frac{b^{1/3}}{2^{3/2}} \paren{\sqrt{1 + 8 \frac{\dflog_2^2}{b^{2/3}}} - 1}
\end{equation}
and the center-of-mass velocity at the peak $\pctwotot$ is
\begin{equation}
    \label{eq_p2tot_soln}
    \pctwotot = \frac{\pctworel}{2}\paren{\sqrt{1 + 4 \frac{\dflog_2^2}{\pctworel^2}} - 1} .
\end{equation}

Fortunately, for collisions between $f_1$ and $f_1$, symmetry allows us to conclude that $\ptot = 0$ at the Gamow peak. The relative velocity at the shifted peak $\pconerel \doteq \sqrt{2} \pcone$ is given by the largest real root of the polynomial
\begin{equation}
    \label{eq_p1rel_soln}
    y_1(s) \doteq s^3 - \dflog_1 s - b .
\end{equation}

The Appendix provides more details about the derivation of these expressions for the peak shift. Using \eqref{eq_G_fraction}, these results yield a corrected utility function
\begin{widetext}
    \begin{equation}
        \begin{split}
        \label{eq_G_k_corrected}
        G(k) = &\frac{S(\pctworel)}{b^{1/3}S(b^{1/3})}\frac{\pctwotot (\pctworel + \pctwotot)^2}{\dflog_2}\sqrt{\frac{3}{\chi_\mr{c2,rel}''\chi_\mr{c2,tot}''}}\frac{e^{-\chi_\mr{c2}}}{e^{-\frac{3}{2}b^{2/3}}} 
        \\& \times  \Bigg[  \pctwo^2 \frac{2\gamma_\mr{c2}^3 - 3(\gamma_\mr{c2}^2 + 1)\tan^{-1}(\gamma_\mr{c2}) + 3\gamma_\mr{c2}}{3\gamma_\mr{c2}^3} 
         - 2\paren{1 - \frac{\tan^{-1}(\gamma_\mr{c2})}{\gamma_\mr{c2}}} 
        - g_\mr{c2} \frac{6(\gamma_\mr{c2}^2 + 2)\tan^{-1}(\gamma_\mr{c2}) - 2\gamma_\mr{c2}(\gamma_\mr{c2}^2 + 6)}{3\gamma_\mr{c2}^3}\Bigg]
        \\& + \frac{\pconerel S(\pconerel)}{b^{1/3}S(b^{1/3})}\sqrt{\frac{3}{\chi_\mr{c1}''}} \frac{e^{-\chi_\mr{c1}}}{e^{-\frac{3}{2}b^{2/3}}} \Bigg[  \pcone^2 \frac{3(\gamma_\mr{c1}^2 + 2)\tan^{-1}(\gamma_\mr{c1}) - \gamma_\mr{c1}(\gamma_\mr{c1}^2 + 6)}{3\gamma_\mr{c1}^3} \Bigg] .
        \end{split}
    \end{equation}
\end{widetext}
This formula \eqref{eq_G_k_corrected} may be viewed as the central result of this work. We have defined
\begin{eqnarray}
    \label{eq_chi_c2_def}
    \chi_\mr{c2} &\doteq& \frac{b}{\pctworel} + \half \pctworel^2 + \half \pctwotot^2 , \\
    \label{eq_chi_c2_doubleprime_rel_def}
    \chi_\mr{c2,rel}'' &\doteq& \frac{2b}{\pctworel^3} + 1 + \frac{\dflog_2}{(\pctworel + \pctwotot)^2}, \\
    \label{eq_chi_c2_doubleprime_tot_def}
    \chi_\mr{c2,tot}'' &\doteq& 1 + \frac{\dflog_2}{(\pctworel + \pctwotot)^2} ,
\end{eqnarray}
and $\gamma_\mr{c2}$ and $g_c$ are given by \eqref{eq_gamma_def} and \eqref{eq_g_def}, respectively, evaluated at $\pctwo$, while
\begin{eqnarray}
    \label{eq_chi_c1_def}
    \chi_\mr{c1} &\doteq& \frac{b}{\pconerel} + \half \pconerel^2 , \\
    \label{eq_chi_c1_doubleprime_def}
    \chi_\mr{c1}'' &\doteq& \frac{2b}{\pconerel^3} + 1 - \frac{\dflog_1}{\pconerel^2} ,
\end{eqnarray}
and $\gamma_\mr{c1}$ is evaluated at $\pcone$.

Although complicated, \eqref{eq_G_k_corrected} is an explicit function of the wavenumber and temperature. Along with \eqref{eq_G_def} and a known turbulent energy spectrum, $G$ allows the reactivity of an arbitrary subsonic flow field to be calculated. The accuracy of this calculation is the subject of the following section. 
The much simpler asymptotic formula \eqref{eq_G_k_asymptotic} is recovered in the limit where $\dflog_1$ and $\dflog_2$ are both small compared to $b^{2/3}$. Although it systematically underestimates $\Phi$ under laboratory fusion-relevant conditions, the simplicity of \eqref{eq_G_k_asymptotic} makes it useful for back-of-the-envelope estimates.

\begin{figure}
    \centering
    \includegraphics[width=0.95\columnwidth]{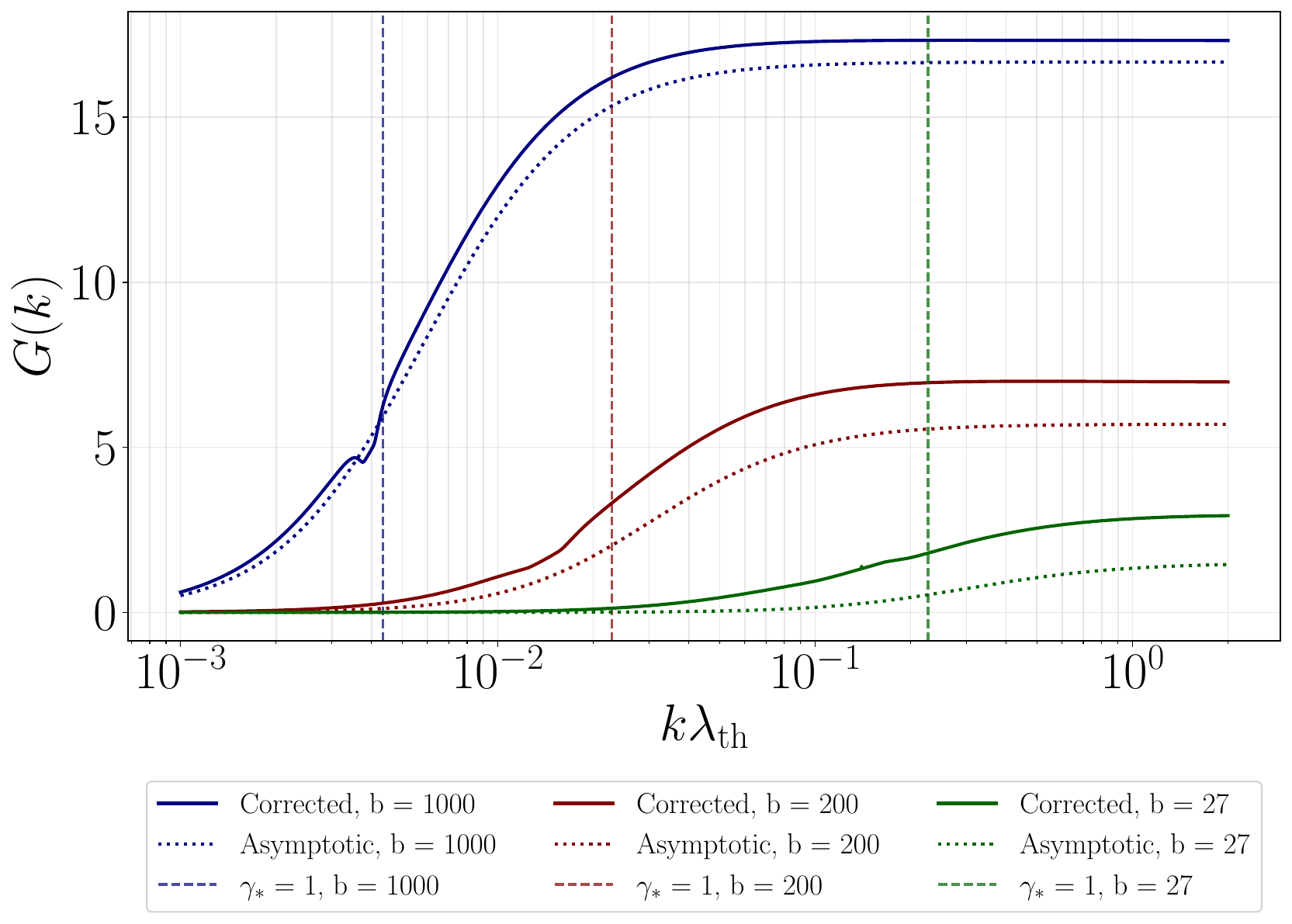}
    \caption{Utility function $G(k)$ as a function of the wavenumber $k$ normalized to the thermal mean free path. Dotted lines show the corrected result \eqref{eq_G_k_corrected} and solid lines show the asymptotic result \eqref{eq_G_k_asymptotic}. Vertical dashed lines show the location where $\gamma_* = 1$. The $b=27$ curve corresponds roughly to DD reactions at $3~\text{keV}$.}
    \label{fig_G_comparison_k_b}
\end{figure}

The asymptotic and corrected results for $G(k)$ are compared in Fig.~\ref{fig_G_comparison_k_b} for a variety of $b$ values. The $b=27$ curve corresponds to a DD plasma at $3~\text{keV}$. The other curves are not directly relevant to a DD fusion reactor because they correspond to sub-keV temperatures where very little fusion occurs, but they are useful for understanding the scaling of $G$ and may be relevant to other fusion reactions. The small dip in the $b=1000$ curve is simply an artifact of the approximation method used for finding the shift in the Gamow peak. 

Clearly, the reactivity enhancement is largest for large $b$ (low temperature). The close agreement between the asymptotic and corrected results for $b=1000$ underscores the observation that \eqref{eq_G_k_asymptotic} is asymptotically valid, albeit at unrealistically low temperatures. The wavenumbers $k_*$ where $\gamma_* = 1$ are marked by vertical dashed lines for each $b$ value. Heuristically, $k_*$ lies near the inflection point of each curve, meaning that, while going to larger $k$ further increases $G(k)$, there are diminishing returns to putting energy in smaller eddies. For large $k$, as predicted by the arguments in \S\ref{sec_asymp}, $G(k)$ asymptotes to a constant.

%In summary, \eqref{eq_G_k_corrected} should be viewed as the central analytical of this work. Because \eqref{eq_G_k_asymptotic} and \eqref{eq_G_k_corrected} exhibit similar 

\section{Numerical results}
\label{sec_numerics}

This section describes the numerical methods used to generate a turbulent flow field, simulate the fast-ion distribution function in that flow and to compute its reactivity. The numerical results are then compared with the predictions of the analytical formula \eqref{eq_G_k_corrected}.

%The analytical expressions derived so far in this work relied on relatively stringent assumptions about the magnitudes of $\hat u$ and $b$. 
%Here, the analytical results obtained above are validated numerically. First, simulations using \eqref{eq_vlasov_v} are compared to the results in \S\ref{subsec_pert_dist}. Next, \eqref{eq_reactivity_def} is solved numerically, both for analytical distributions and for those obtained from simulations, and compared to the estimates in \S\ref{subsec_analytics_reactivity}.

\subsection{Kinetic simulation}
\label{subsec_numerics_dist}

%\subsubsection{Numerical method}

We simulate the kinetic transport of fast particles using a method similar to that described by \FF. As in this rest of this work, we assume that $\kn \ll 1$, meaning that a fluid description is appropriate for the bulk of the plasma. Therefore, for thermal particles, $f(\bs w)$ is close to $f_M(\bs w)$ everywhere. A flow field $\bs u(\bs x)$ is prescribed on a periodic domain. Unlike in \S\ref{sec_analytics}, the Mach number is no longer required to be small. 
We solve \eqref{eq_vlasov_v} using a modified BGK operator \eqref{eq_bgk_def} with collision frequency given by
\begin{equation}
    \label{eq_nu_numerical}
    \nu(p) = \frac{\nu_0}{p^3} \left[\frac{\mr{erf}\paren{\frac{p}{\sqrt{2}}} - \sqrt{\frac{2}{\pi}}pe^{-\half p^2}}{\frac{1}{4}p^2}\right] + \nu_0 \sqrt{\frac{m_e}{m}},
\end{equation}
where $\mr{erf}(x)$ is the error function. 
%Note that ${\nu(p \gg 1) \sim \nu_0/p^3}$, as we asserted in \S\ref{sec_intro}. 
Because $\nu$ is not a constant, this operator does not conserve overall density, momentum, or energy. However, since we are only interested in the behavior of tail particles, this is not a problem; the result is simply normalized at the end of the simulation to enforce the conservation of particles. 

The simulation is carried out on a discretized phase space in a periodic box with 160 points per spatial dimension. The velocity grid spacing is $\Delta v \approx 0.33 \vth $. The distribution function is found by initializing $f$ as Maxwellian everywhere, with drift velocities given by the local $\bs u$, and iterating forward along trajectories, making at least twenty passes through the box to ensure that the distribution function approaches its local steady state. Adjusting the grid spacing and number of passes are verified to have a negligible effect on the final distribution.

\begin{figure}
    \centering
    \includegraphics[width=0.95\columnwidth]{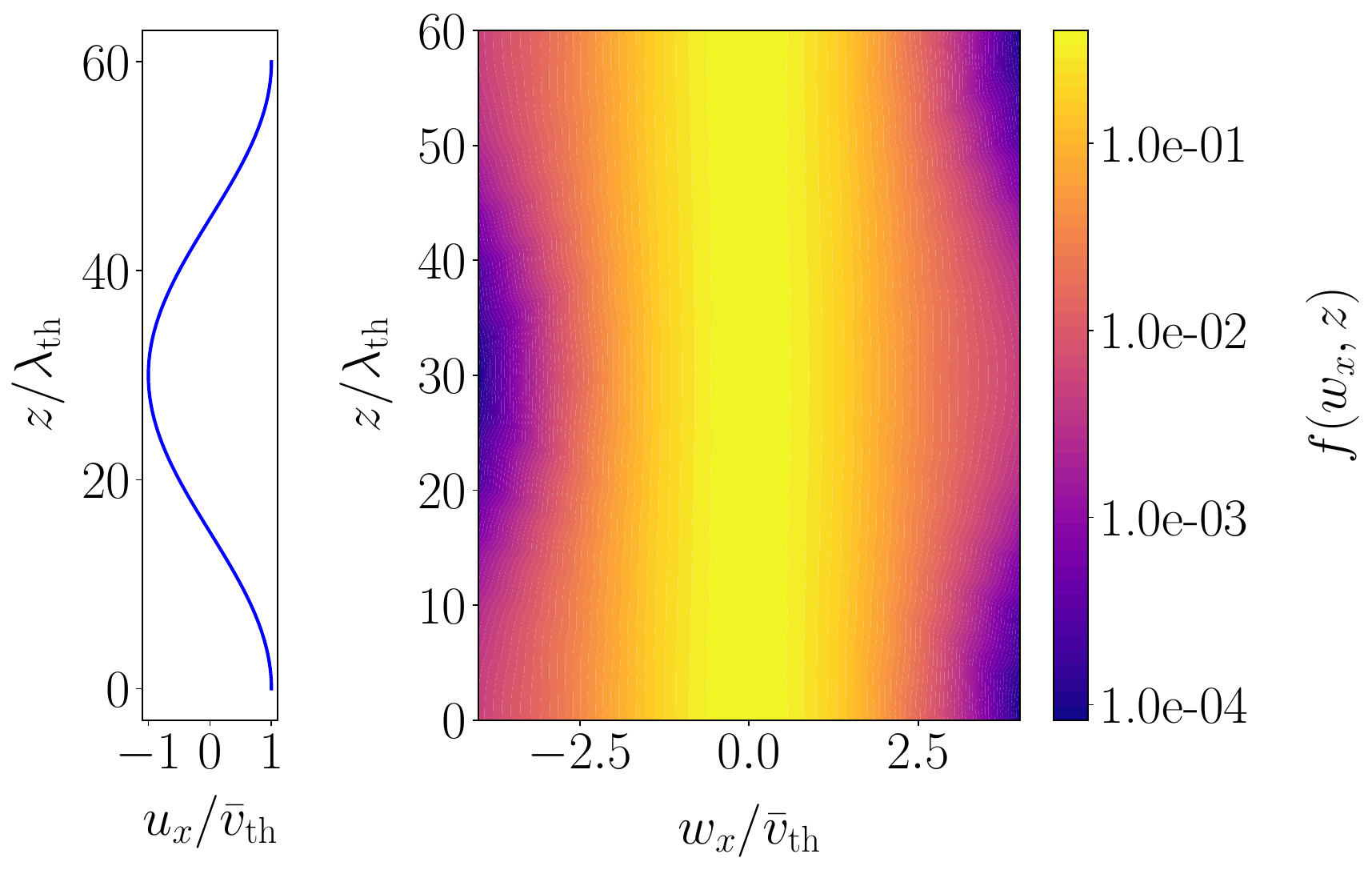}
    \caption{Simulated distribution function in the $z-w_x$ plane for a sinusoidal shear flow with $u_0 = \vth$ and $k = 1/10\lambda_\mr{th}$. The left panel shows the background flow profile $u_x(z)$. Normalization of $f$ is arbitrary.}
    \label{fig_sin_shear}
\end{figure}

The effect of flows on $f$ is most easily seen in a planar shear flow. 
For clarity, we show in Fig.\ref{fig_sin_shear} the numerically computed distribution function for a flow given by 
\begin{equation}
    \bs u(\bs x) = u_0 \bs e_x \sin(k z) ,
\end{equation}
where $z = \bs x \cdot \bs e_z$. For the simulation shown, $u_0 = \vth$ and $k\lambda_\mr{th} = 2\pi/60$ so that one period of the flow fits in the simulation domain. The distribution function is shown projected onto the $(w_x, z)$ plane, integrating over the other coordinates. Since $f$ is shown in terms of the peculiar velocity $\bs w$, a distribution in perfect local thermal equilibrium (at uniform temperature and density) would appear not to vary in $z$. Indeed, this is true in the highly collisional region near $v = \vth$, manifesting in Fig.~\ref{fig_sin_shear} as the streak of high phase-space density around $w_x = 0$, which is essentially constant in $z$.

For faster ions, on the other hand, the flow shear leads to non-Maxwellian behavior. Consider, for example, an ion initially sampled at $w_x = 3\vth$ at $z=0$ and with $w_z > 0$. As this ion travels upward, it moves into a region where $\bs u$ points toward $-\bs e_x$. Assuming that this ion has not yet collided, its will now be at larger $w_x$ than it was before. This can be seen in Fig.\ref{fig_sin_shear} as a bump in phase-space density at large $w_x$ around $z=30$, the region of largest negative velocity; of course, the same effect applies in reverse, enhancing the $w_x < 0$ region near $z=0$.

\subsection{Computation of reactivity}
\label{subsec_numerics_reactivity}

Fusion reactivity can be found from the simulated distribution function according to \eqref{eq_reactivity_def}. Because phase space is already discretized, this can be done by simple matrix multiplication. At each spatial location, the velocity grid is flattened to a one-dimensional vector $F$ such that
\begin{equation}
    F_i \doteq f(\bs W_i) ,
\end{equation}
where $W_i$ is the velocity grid at the $i$th point in flattened velocity space. 
A matrix $\widehat \Sigma$ is precomputed with the value of $\sigma v$ for each pair of velocities, \textit{viz.}
\begin{equation}
    \widehat \Sigma_{ij} \doteq \frac{\Delta v^6}{2} \sigma\paren{\frac{|\bs W_i - \bs W_j|}{\sqrt{2}\vth}} |\bs W_i - \bs W_j| .
\end{equation}
The reactivity is then given by 
\begin{equation}
    \label{eq_Gamma_F_Sigma}
    \reactop[F, F] = F^\mr{T} \widehat \Sigma F .
\end{equation}
Although computationally intensive, this method is feasible for the numerical grids used in this work. It was verified that increasing the grid resolution does not substantially affect the result for the distributions and cross sections considered here. 
The DD cross section of Bosch and Hale \cite{Bosch_Hale_1992}, which represents the S factor by a Padé approximant, is used to compute $\widehat \Sigma$.

%Two cross section models are used in this work. For DD reactivity, we use the standard Bosch-Hale cross section \cite{Bosch_Hale_1992}, which represents the S-factor by a Padé approximant. For simplicity, we additionally use the simplified cross section given by \eqref{eq_sigma_S} with $b = 44.4$, corresponding to the Gamow energy for DD fusion. While less practically useful, this simplified model allows for a more straightforward comparison with the analytical results derived in \S\ref{sec_analytics}. It is thereby easier to determine whether deviations from the analytical result are due to approximations made in the kinetic theory, e.g. \eqref{eq_vlasov_f1} and approximation of the reactivity integral, e.g. \eqref{eq_reactivity_maxwellian_leading}, as opposed to features of the DD cross section.

\subsection{Background flows}
\label{subsec_numerics_background_flows}

The fundamental question at hand in this work is how fusion reactivity behaves in turbulent plasmas in general. Ideally, to compute the reactivity in some system of interest, the fluid flow would be simulated or measured with extremely high precisision. In practice, however, flow profiles are rarely known to the necessary precision. 
In this work, therefore, we consider an ensemble of flow fields with widely varying properties to assess the accuracy of \eqref{eq_G_k_corrected} and to benchmark the range of reactivity enhancements that can be expected for a given set of macroscopic parameters.
%Naturally, the corresponding reactivity enhancements also vary. Nonetheless, computing $\Phi$ for each flow allows determination of the range of reactivity enhancements that might be expected based on a few known macroscopic parameters. Comparison with the analytical results obtained in \S\ref{sec_analytics} allows the accuracy of \eqref{eq_G_k_corrected} to be assessed.

In the periodic box used for simulations, any flow field can be represented as a series of Fourier modes. Each mode is described by a triplet of nonnegative integers ($n_x, n_y, n_z$) such that the wavevector $\bs k$ is 
\begin{equation}
    \label{eq_k_num_def}
    \bs k = \frac{2\pi n_x}{L_x} \bs e_x + \frac{2\pi n_y}{L_y} \bs e_y + \frac{2\pi n_z}{L_z} \bs e_z ,
\end{equation}
where $L_x$, $L_y$, and $L_z$ are the lengths of the periodic box in each direction. We consider only modes whose polarization vector $\widetilde{\bs u}$ is perpendicular to $\bs k$, ensuring that the flow is solenoidal. 
In the following cases, two-dimensional flows ($n_y = 0$) are used to speed the computation (this method was benchmarked against fully three-dimensional flows to verify that the results are not sensitive to this simplification). 
Flows are generated as follows. For a given box size, a set of modes ${(n_x, n_y, n_z) = (0, 0, 1)}$ to $(n_x, n_y, n_z) = (20, 0, 20)$ is generated with random phases. The amplitude of each mode is chosen randomly from a uniform distribution $[u_k/5, u_k]$ where $u_k = u_0 (n_x^2 + n_z^2)^{-2/3}$ and $u_0$ is a constant. The result approximates a Kolmogorov cascade. The ensemble of flows is generated by scanning values of the overall flow magnitude $u_0$ from $0.2\vth$ to $\vth$; two flows are generated for each value of $u_0$. 

\begin{figure}
    \centering
    \includegraphics[width=0.95\columnwidth]{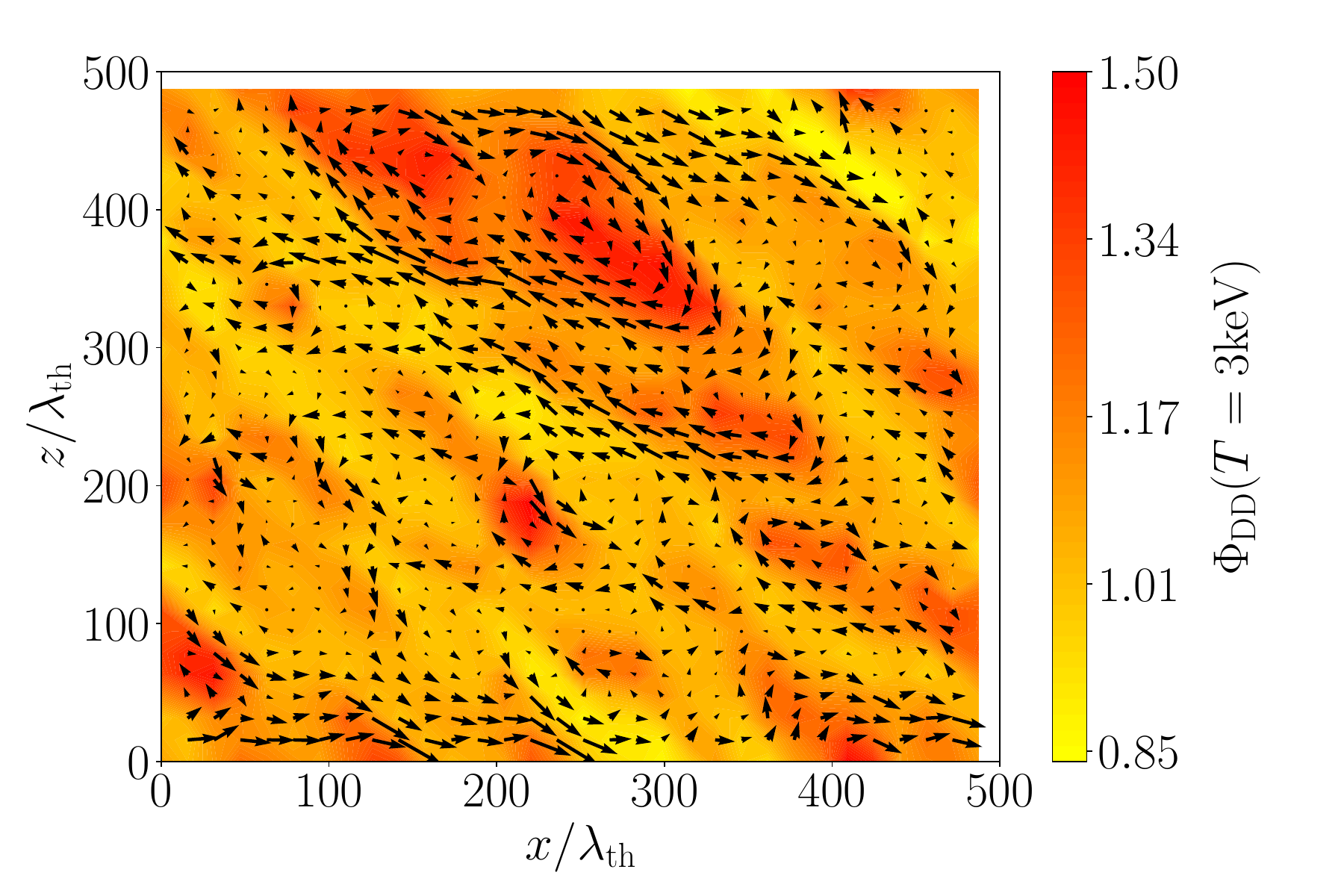}
    \caption{Reactivity enhancement in a 2D turbulent flow field. Black arrows show the flow direction.}
    \label{fig_2d_turb}
\end{figure}

For a flow field generated in this way with $\avg{u^2} = 0.13\vth^2$, the spatial structure of the reactivity enhancement is shown in Fig.~\ref{fig_2d_turb}. The patches of $\Phi > 1$, and a few regions of $\Phi < 1$, stand in sharp contrast to the conventional local picture of reactivity, where $\Phi = 1$ everywhere by definition (in an isothermal system like this one). In the critical early stages of a fusion burn, the appearance of local spots of increased reactivity may be even more significant than the global average reactivity.

\subsection{Comparison to analytical result}

\begin{figure} 
    \centering
    \begin{subfigure}[t]{\columnwidth}
        \centering
        \includegraphics[width=\textwidth]{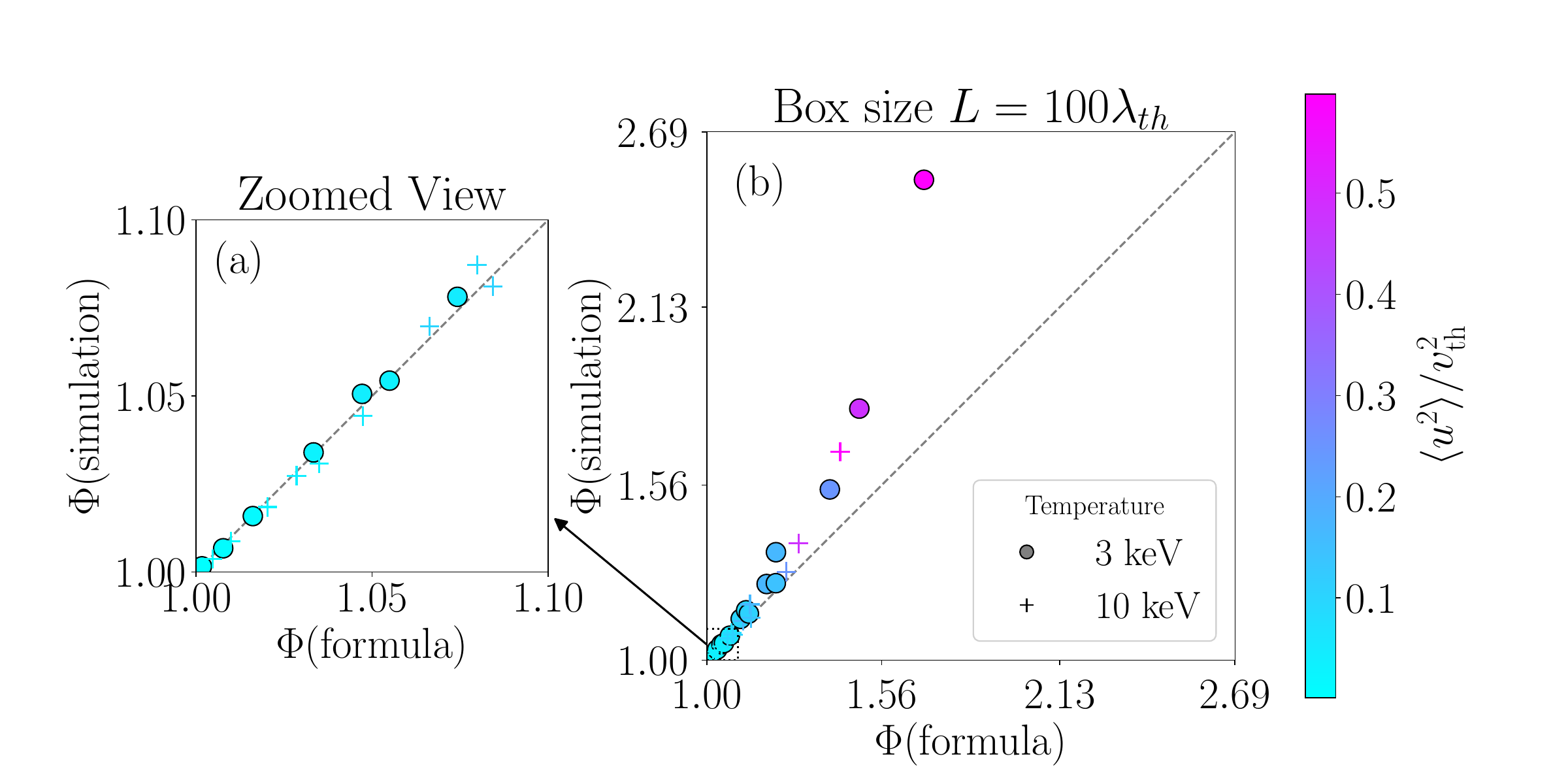}
    \end{subfigure}
    \hfill
    \begin{subfigure}[t]{\columnwidth}
        \centering
        \includegraphics[width=\textwidth]{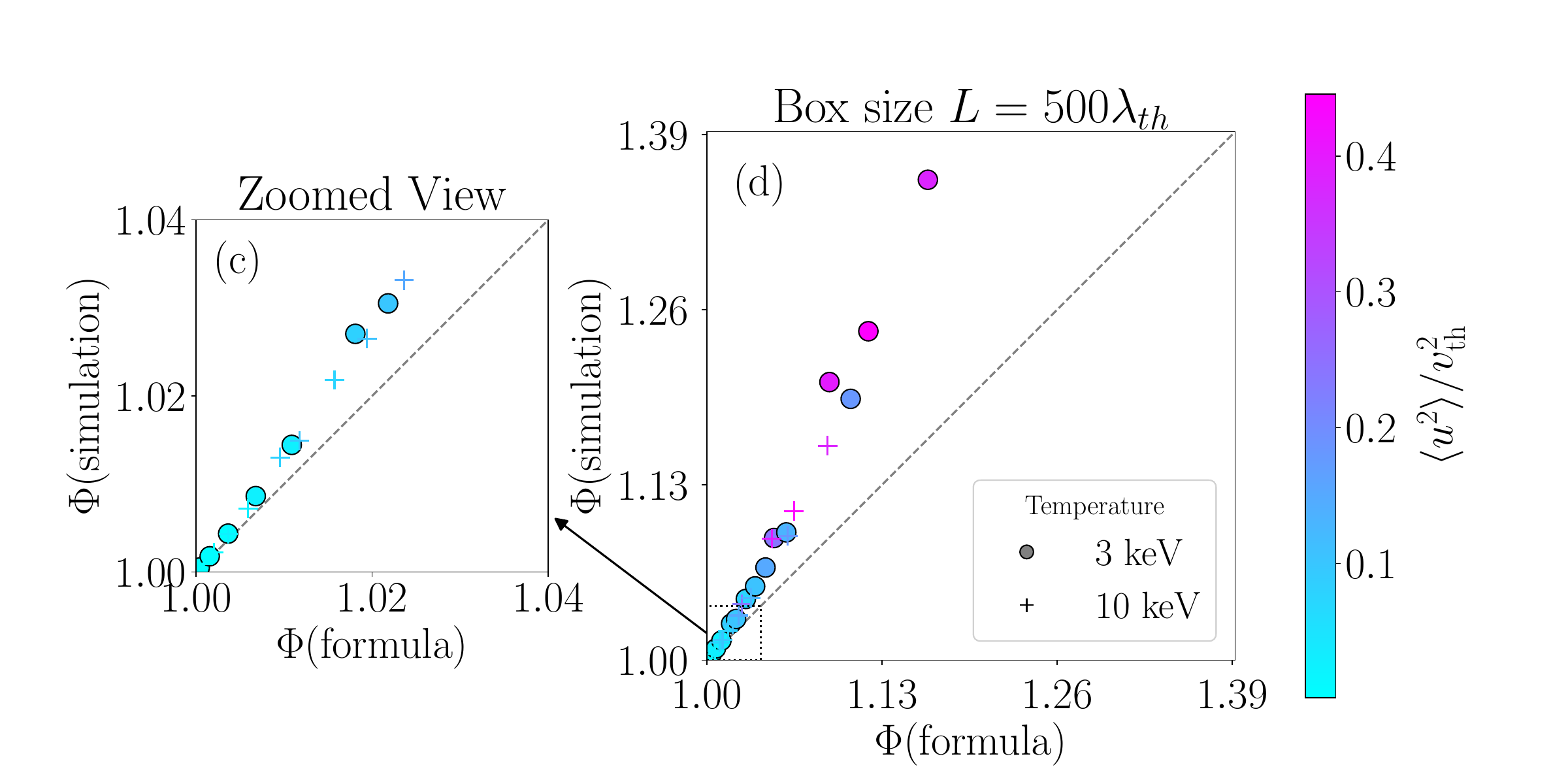}
    \end{subfigure}
    \caption{Reactivity enhancement for DD fusion using Bosch-Hale cross sections. Turbulent flows are generated in boxes of size $L = 100\lambda_\mr{th}$ (a,b) and $L = 500\lambda_\mr{th}$ (c,d). $\avg{u^2}$ is the mean square flow velocity. On the horizontal axis, $\Phi$ is obtained from \eqref{eq_G_k_corrected} and on the vertical axis, $\Phi$ is obtained numerically. The left panel of each plot highlights the low Mach-number regime, in which the assumptions of the analytical theory are valid.}
    \label{fig_reactivity_comp}
\end{figure}

For an ensemble of flow fields generated as described in \S\ref{subsec_numerics_background_flows}, the numerically computed reactivity is compared with the analytical predictions of \eqref{eq_G_k_corrected} in Fig.~\ref{fig_reactivity_comp}. The formula for $\nu(w)$ given in \eqref{eq_nu_numerical} is used for both numerical and analytical calculations. 
Flows are generated in two-dimensional periodic boxes of size $L = 100\lambda_\mr{th}$ and $L = 500\lambda_\mr{th}$ (where $L = L_x = L_z$) and reactivity is evaluated at $3~\mr{keV}$ and $10~\mr{keV}$. For flows with low Mach number, theory is in excellent agreement with simulation. At larger Mach number, theory systematically underpredicts the reactivity enhancement. 
It is to be expected that the analytical theory would break down at large Mach number, where the assumption that $u \ll \vth$ is not valid. 
Agreement is better in the smaller box (meaning flows on shorter length scales), which can be attributed to the scheme described in \S\ref{subsec_analytics_reactivity} for estimating the shift in the Gamow peak. For smaller $\gamma$, the shift in the peak location is greater and \eqref{eq_G_k_corrected} tends to be an underestimate. 
It is interesting to note that the theoretical prediction is nearly always an underestimate. With this in mind, reactivity found using \eqref{eq_G_k_corrected} can be viewed as a conservative approximation. 
The reactivity enhancement factor is larger at low temperature; this scaling appears in all reduced models of the \sfe~discussed above and is easily seen in \eqref{eq_Phi_gk_infty_low_Mach} and \eqref{eq_G_k_asymptotic}. 
Of course, in absolute terms, reactivity generally increases with temperature, so the boost to fusion yield due to TKE does not necessarily decrease with temperature.

\section{Discussion}
\label{sec_discussion} 

The formulas derived in this work, namely \eqref{eq_G_k_corrected} used in combination with \eqref{eq_G_def}, allow the reactivity enhancement to be assessed based on relatively limited information about a system. 
This is important because, in practice, the detailed structure of flows in an ICF target is often diagnostically inaccessible. The results of this work allow $\Phi$ to be estimated if even coarse data about the turbulent energy spectrum can be obtained from observations or simulations. The fact that the \sfe~was only recently identified suggests that $\Phi$ has likely been close to unity in most previous experiments. Estimates in \FF~support this conclusion\citeFF. 
The estimation can be done in reverse as well: precise measurements of fusion yield, along with a detailed quantitative model of the \sfe, could be used to infer information about the turbulent energy spectrum in an experiment. This could have diagnostic utility, even in cases where the overall effect on the experiment is minor.

No ICF experiment has yet been designed with the \sfe~in mind. With deliberate driving of small-scale turbulence, reactivity could be increased much further. 
To give a sense of the plausible size of the effect, Table~\ref{tab_scenarios_parms} considers three scenarios broadly representative of ICF parameter regimes. 
Note that, for simplicity, the derivations in this work considered only reactants of equal mass and charge; for practical purposes, this means DD reactions. However, the results can be taken as a reasonable approximation for reactants with similar masses. Therefore, the following calculation considers DT reactions, treating the DT ions as a single species with mass $m = 2.5m_p$ ($m_p$ is the proton mass) and using the Bosch-Hale cross section\cite{Bosch_Hale_1992} for DT fusion to compute the reactivity.

For each scenario, $\Phi$ is calculated as follows.
For a given density and temperature, it is assumed that TKE is driven at some scale $L_0$ and cascades to smaller scales, where it eventually dissipates. For a given dissipation time scale $t_\eta$, the dissipation length scale can be estimated as $L_\eta = \sqrt{\eta t_\eta}$, where $\eta$ is the kinematic viscosity. For simplicity, the turbulent spectrum is assumed to obey a $-5/3$ power law between $L_0$ and $L_\eta$ with no TKE above or below this range. The energy of the flow is reported as the ratio $\mc T$ between TKE and thermal energy defined as
\begin{equation}
    \label{eq_TKE_ratio_def}
    \mc T \doteq \frac{\mr{TKE}}{3 T} ,
\end{equation}
where TKE is given by \eqref{eq_TKE_spec}. The factor of 3 in the denominator comes from the fact that electrons and ions each have thermal energy density $3nT/2$. This model of the turbulent spectrum is heavily simplified, but it captures the essential parameter dependence of the effect: turbulence is driven on some outer scale and vanishes below some dissipation scale due to viscosity; both scales matter in the \sfe. The reactivity enhancement is calculated using the turbulent energy spectrum along with \eqref{eq_G_def} and \eqref{eq_G_k_corrected}. The previous comparison to numerical results (Fig.~\ref{fig_reactivity_comp}) suggests that the figures in Table~\ref{tab_scenarios_parms} are most likely underestimates; in reality, a smaller $\mc T$ would probably be sufficient to achieve the same $\Phi$.

The first scenario considered here corresponds to a laser indirect-drive experiment, for example on the National Ignition Facility (NIF). The second scenario corresponds to a fast-ignition experiment, where fuel is compressed to high density and then rapidly heated by a fast ignitor pulse; the parameters in Table~\ref{tab_scenarios_parms} represent the hot spot immediately after heating, before the hot material has had time to expand against the surrounding cold fuel. The third scenario corresponds to a z pinch at conditions well below ignition but relevant to experiments. 
The specific parameters listed for each scenario are rough figures taken to represent each regime, rather than modeling a particular experiment. 
It is assumed that a significant amount of TKE is driven in each scenario. While the high temperature of the fuel means that the TKE will dissipate quickly, it need only persist long enough to help the hot spot to ignite. For the indirect-drive and z-pinch scenarios, $t_\eta$ is chosen to be long enough to survive the final stage of compression, in which the fuel becomes highly viscous. 

The fast-ignition scenario is entirely different. Prior to the ignitor pulse, the compressed fuel is cold and therefore relatively inviscid.
The time scale on which TKE needs to persist in hot, viscous fuel is therefore simply the heating time, rather than the compression time; the former time scale is significantly shorter in fast ignition. 
It follows that turbulence can be driven on smaller scales without being lost to viscous dissipation and, as a result, larger reactivity enhancements are possible. As discussed in \FF, this has profound implications for the design of fast-ignition experiments. Fuel ignites at a lower temperature when $\Phi > 1$. Charged fusion products stop more quickly in colder fuel, so the hot spot can be made smaller. Therefore, in addition to the energy saved by heating the fuel to a lower temperature, an even larger amount of energy is saved because the ignitor can heat a much smaller fuel mass and still reach ignition\citeFF.

\begin{table}
    \centering
    \renewcommand{\arraystretch}{2} % Increase row height
    \begin{tabular}{|>{\centering\arraybackslash}p{0.3\columnwidth}|>{\centering\arraybackslash}p{0.2\columnwidth}|>{\centering\arraybackslash}p{0.2\columnwidth}|>{\centering\arraybackslash}p{0.2\columnwidth}|}
        \hline
        \diagbox[width=0.3\columnwidth,height=3.2\baselineskip,trim=l,innerleftsep=0pt,innerrightsep=0pt,linewidth=0.75pt]{Parameter}{Scenario} & Indirect Drive & Fast \newline Ignition & Z Pinch \\
        \hline
        Density $\rho$ & 40 g/cm$^3$ &  300 g/cm$^3$& 5 g/cm$^3$\\ \hline
        Temperature $T$ & 10 keV & 8 keV & 3 keV \\ \hline
        TKE Ratio $\mc T$ & 1/3 & 1/3 & 2/3 \\ \hline
        Forcing Scale $L_0$ & 10 $\mu$m& 1 $\mu$m& 50 $\mu$m\\
        \hline
        Dissipation Time $t_\eta$& 1 ns & 0.1 ns & 5 ns \\ \hline \hline
        Dissipation Scale $L_\eta$ & 7.8 $\mu$m & 0.1 $\mu$m & 11 $\mu$m \\ \hline
        Gamow mean free path $\lambda_*$ & 23 $\mu$m & 3.0 $\mu$m & 26 $\mu$m \\ \hline
        Reactivity Enhancement $\Phi$ & 1.6 & 3.1 & 1.4 \\ \hline
    \end{tabular}
    \caption{Enhancements of DT reactivity calculated using \eqref{eq_G_k_corrected} for three generic ICF scenarios. The first five rows are fixed parameters and the final three rows are computed as described in the text.}
    \label{tab_scenarios_parms}
\end{table}

It is clear from Table~\ref{tab_scenarios_parms} that a significant increase in reactivity is possible on the time and length scales relevant to ICF, provided that large amounts of TKE can be driven in the fuel. 
In fact, high levels of TKE are a common feature of ICF experiments. In z pinches, flow kinetic energy has been observed to exceed thermal energy by a factor of several times in some cases \cite{Giuliani_et_2014,Kroupp_et_2011,Maron_2020}. Detailed energy accounting suggests that much of this TKE is contained in turbulent eddies \cite{Kroupp_et_2018,Davidovits_Kroupp_Stambulchik_Maron_2021}. 
In laser indirect-drive ICF, high fusion yield has generally been associated with decreasing TKE \cite{Ralph_et_2024,Pak_et_2024,Ma_et_2013,Haines_Grinstein_Fincke_2014}. Nevertheless, the kinetic energy contained in flows is often substantial \cite{Weber_Clark_Cook_Busby_Robey_2014,Weber_et_2015,Clark_et_2015,Kritcher_et_2024}. 
Viscosity, of course, dissipates small-scale turbulent structures, but several processes drive new flows at late times. Imperfections in the target, for example due to the fill tube or support tent, can launch jets of material into the hot spot at high speed, producing large shear \cite{Weber_et_2015,Haines_et_2020,Pak_et_2024}. Perturbations at interfaces can grow by the Rayleigh-Taylor (RT) or Richtmyer-Meshkov (RM) instability and drive turbulent flows \cite{Hammel_et_2010,Zhou_Sadler_Hurricane_2025,Thomas_Kares_2012,Grinstein_et_2024}. 

While large TKE is clearly possible in ICF, an important question remains as to whether it can be driven primarily in small (micron-scale) eddies. For large-scale eddies ($k \ll k_*$), which matter little for the \sfe~($G(k) \ll 1$), the standard ICF paradigm applies and such TKE is simply a waste of energy. 
For context, Table~\ref{tab_scenarios_parms} lists the Gamow mean free path $\lambda_* \doteq \lambda(b^{2/3}/\sqrt{2})$ for each scenario ($k_* \doteq 1/\lambda_*$ is the corresponding wavenumber). Recall from \S\ref{sec_analytics} that, at least heuristically, turbulence on a scale $k \sim k_*$ is necessary for a large reactivity enhancement.  In the scenarios listed here, turbulence is driven on scales slighly longer than the Gamow mean free path. 

To drive turbulence on small scales, short-wavelength perturbations could be deliberately seeded at interfaces prone to RT or RM instability. Waves, shocks\cite{Zhang_Davidovits_Fisch_2025}, and particle beams\cite{Malkin_Fisch_2002}, as well as structured materials such as foams, can also drive turbulence during compression\cite{Haines_et_2023,Murphy_et_2021}. 
In addition to being driven on the right scale, turbulence must be driven in the right place. Turbulent mixing is a major loss mechanism \cite{Larroche_Rinderknecht_Rosenberg_2018,Albright_et_2022,Hurricane_et_2024,Murphy_et_2021}, meaning that turbulence at the edges of the hot spot is likely to have a deleterious effect overall, even if it exists on scales relevant to the \sfe. 
Therefore, it is probably desirable that the mechanism driving TKE act primarily near the center of the hot spot and leave a relatively quiescent region on the periphery. 
The fundamental tradeoff here is that large-scale turbulence dissipates more slowly and is likely easier to drive, but small-scale turbulence produces a larger reactivity enhancement.

Clearly, designing an ICF system to benefit from the \sfe~involves a number of technical challenges. 
Such a design calls for the reverse of what decades of ICF development have striven to accomplish: driving turbulence rather than suppressing it (granting that this turbulence must be driven carefully on the right scales and in the right place). Driving the necessary flows requires an understanding of the dynamics of dense, compressing, turbulent plasmas. 
These questions are of fundamental academic interest, but the practical benefits of designing such an ICF system could also be enormous. 
In fast ignition, for example, leveraging the \sfe~enables a potentially order-of-magnitude reduction in ignitor energy. 
It is possible that embracing small-scale turbulence allows some symmetry requirements to be relaxed, leading to more economical target manufacturing. This is far from certain -- large-scale flows and turbulent mixing between hot and cold material still need to be avoided -- but it is an appealing prospect. 
Because the Gamow energy is higher for fuels other than DD and DT, the \sfe~ is expected to be larger for these reactions. An understanding of this effect could therefore be instrumental in the design of ICF systems burning ``advanced'' or aneutronic fuels.

\section*{Acknowledgments}
This work was supported by the Center for Magnetic Acceleration, Compression, and Heating (MACH), part of the U.S. DOE-NNSA Stewardship Science Academic Alliances Program under Cooperative Agreement DE-NA0004148.

\medskip

\appendix

\section{Calculation of the corrected utility function}
\label{appendix_peak_shift}

This appendix fills in details of the derivation of the shift in the location of peak reactivity underlying \eqref{eq_G_k_corrected}. To evaluate $\reactop[f_2, f_0]$, we begin with the integral
\begin{widetext}
\begin{equation}
    \label{eq_app_starting_integral}
    I \doteq \int d^3p \int d^3p' \frac{\sqrt{2}S\paren{\frac{|\bs p - \bs p'|}{\sqrt{2}}}}{|\bs p - \bs p'|} \overline{\varphi}_{20}(p) \frac{e^{-\frac{\sqrt{2}b}{|\bs p - \bs p'|} - \half p^2 - \half {p'}^2}}{(2\pi)^3} .
\end{equation}
Changing variables to $\bs p_\mr{rel} \doteq (\bs p - \bs p')/\sqrt{2}$ and $\bs p_\mr{tot} \doteq (\bs p + \bs p')/\sqrt{2}$ gives
\begin{equation}
    \label{eq_app_I_prel_ptot}
    I = \int d^3\prel d^3\ptot \frac{S(\prel)}{\prel} \overline{\varphi}_{20}\paren{\frac{| \bs p_\mr{rel} + \bs p_\mr{tot}|}{\sqrt{2}}} \frac{e^{-\frac{b}{\prel} - \half \prel^2 - \half \ptot^2}}{(2\pi)^3} .
\end{equation}
Now suppose that $\overline{\varphi}_{20}(p) = C p^{\dflog}$. Let $\bs p_\mr{rel}\cdot \bs p_\mr{tot} \doteq \beta \prel \ptot$. Then integrating over all angles except the angle between $\bs p_\mr{rel}$ and $\bs p_\mr{tot}$, we have 
\begin{equation}
    I = \frac{C}{\pi} \int_0^\infty d\prel \int_{-1}^1 d\beta \int_0^\infty d\ptot \frac{S(\prel)}{\prel} \frac{(\prel^2 + 2\beta\prel\ptot + \ptot^2)^{\dflog/2}}{2^{\dflog/2}} e^{-\frac{b}{\prel} - \half \prel^2 - \half \ptot^2}.
\end{equation}
Evaluating the integral over $\beta$ and assuming that $(\prel + \ptot)^{\dflog} \gg (\prel - \ptot)^{\dflog}$, we have
\begin{equation}
    I \sim \frac{C}{\pi 2^{\dflog/2}} \int_0^\infty d\prel \int_0^\infty d\ptot \frac{S(\prel)}{\prel} \frac{1}{2\prel \ptot (\half \dflog + 1)} (\prel + \ptot)^{\dflog + 2} e^{-\frac{b}{\prel} - \half \prel^2 - \half \ptot^2} .
\end{equation}

To proceed, we move the final term of the prefactor into the exponent, which is defined by
\begin{equation}
    \label{eq_app_chi_def}
    \chi_2 \doteq -\frac{b}{\prel} - \half \prel^2 - \half \ptot^2 + (\dflog + 2)\ln(\prel + \ptot) .
\end{equation}

\end{widetext}
The peak of the integrand is located where $d\chi_2/d\prel = 0$ and $d\chi_2/d\ptot = 0$. Defining $\dflog_2 \doteq \dflog + 2$, the iterative procedure proceeds as follows. First, we find $\ptot$ at the Gamow peak using the initial guess $\prel = b^{1/3}$, as would be true if $\dflog = 0$. The peak is located where
\begin{equation}
    0 = -\ptot + \frac{\dflog_2}{b^{1/3} + \ptot} ,
\end{equation}
from which we obtain
\begin{equation}
    \label{eq_app_ptot_1}
    \ptot = \half \paren{b^{1/3} + \sqrt{b^{2/3} + 4\dflog_2}} .
\end{equation}
Using this value of $\ptot$ and maximizing $\chi_2$ with respect to $\prel$, we find
\begin{equation}
    0 = \frac{b}{\prel^2} - \prel + \frac{\dflog_2}{\prel + \half \paren{b^{1/3} + \sqrt{b^{2/3} + 4\dflog_2}}} .
\end{equation}
Requiring $\pctworel$ to be the largest real solution to this equation, we recover \eqref{eq_p2rel_soln}. Finally, using \eqref{eq_app_ptot_1} and replacing $b^{1/3}$ with $\pctworel$ gives the expression for $\pctwotot$ in \eqref{eq_p2tot_soln}. This iterative procedure could be continued, but the results obtained at this stage are reasonably accurate. For higher accuracy, numerical integration is more appropriate.

Evaluating \eqref{eq_app_starting_integral} by expanding around the peak at $\prel = \pctworel$ and $\ptot = \pctwotot$ leads to the first two lines of \eqref{eq_G_k_corrected}. The third line, which corresponds to $\reactop[f_1, f_1]$, is obtained through a nearly identical calculation but with a simpler calculation because symmetry allows us to conclude that $\ptot = 0$ at the Gamow peak.

\begin{figure}
    \centering
    \includegraphics[width=0.8\columnwidth]{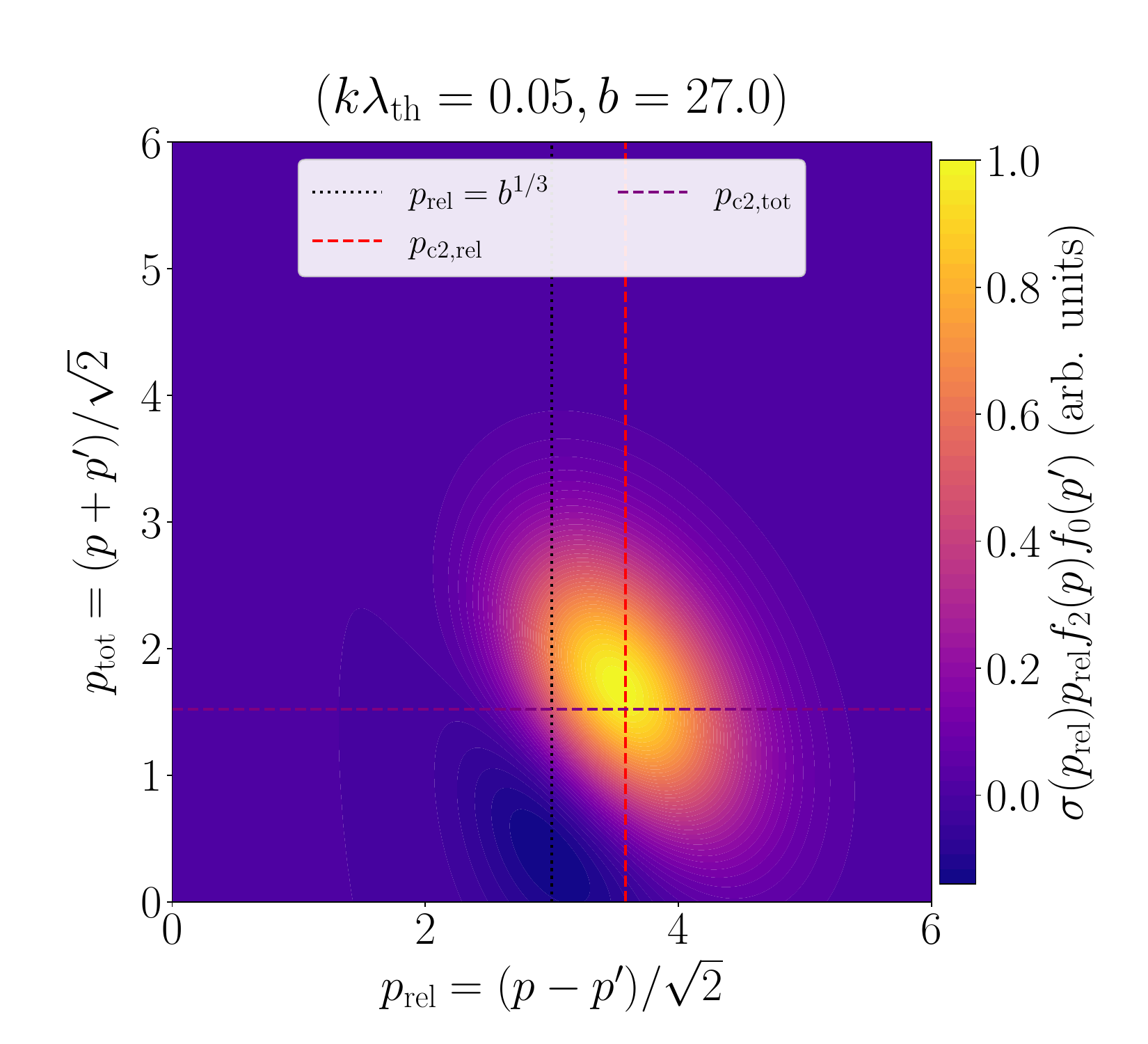}
    \caption{Rapidly varying part of the integrand in \eqref{eq_app_starting_integral} as a function of $\prel$ and $\ptot$ for a mode with $k = 0.05$ and with $b=27$. The unshifted value of $\prel$ is shown in black and $\pctworel$ and $\pctwotot$ are shown in red and purple respectively. Note that the unshifted peak value of $\ptot$ is zero. Normalization of the colormap is arbitrary.}
    \label{fig_reactivity_integral}
\end{figure}

The shift in the peak location can be seen graphically in Fig.~\ref{fig_reactivity_integral}. Clearly, the approximation scheme derived here does not perfectly recover the peak location but works significantly better than the naive estimate of $\prel = b^{1/3}$. 

%In existing ICF experiments, where the detailed structure of flows in the fuel may be diagnostically inaccessible, but bounds may be placed on the level of TKE, $\Phi$ can be estimated and bounded. 
%In the design of future ICF experiments, these formulas indicate the conditions under which TKE can be expected to produce a beneficial effect.

\bibliography{turbulent_reactivity}

\end{document}

%% file: macros.tex
\newcommand{\bs}[1]{\boldsymbol{#1}}

\newcommand{\mc}[1]{\mathcal{#1}}

\newcommand{\half}[0]{\frac{1}{2}}

\newcommand{\paren}[1]{\left(#1\right)}

\newcommand{\wt}{\widetilde}

\newcommand{\avg}[1]{\langle #1 \rangle}

\newcommand{\mr}[1]{\mathrm{#1}}

\newcommand{\vth}{v_\mr{th}}

\newcommand{\lth}{\lambda_\mr{th}}

\newcommand{\kn}[0]{\mathrm{Kn}}
\newcommand{\gk}[0]{\mathrm{Gk}}

\newcommand{\sfe}{SFRE}